\documentclass[aps,preprint,nofootinbib,eqsecnum]{revtex4}
\usepackage{amsmath}
\usepackage{amsfonts}
\usepackage{amssymb}
\usepackage{graphicx}
\usepackage{amscd}

\begin{document}

\noindent {\small USC-HEP/0406IB1\hfill \hfill hep-th/0407239 \newline
}

{\small \hfill }

{\vskip0.5cm}

\begin{center}
{\Large \textbf{Twistor Superstring in 2T-Physics}}{\footnote{%
This research was partially supported by the US. Department of Energy under
grant number DE-FG03-84ER40168.}}{\Large \textbf{\ }}

\bigskip

{\vskip0.5cm}

\textbf{Itzhak Bars}

{\vskip0.5cm}

\textbf{Department of Physics and Astronomy}

\textbf{University of Southern California}

\textbf{\ Los Angeles, CA 90089-2535, USA}

{\vskip0.5cm}

\textbf{Abstract}
\end{center}

By utilizing the gauge symmetries of Two-Time Physics (2T-physics), a
superstring with linearly realized global SU$\left( 2,2|4\right) $
supersymmetry in 4+2 dimensions (plus internal degrees of freedom) is
constructed. It is shown that the dynamics of the Witten-Berkovits twistor
superstring in 3+1 dimensions emerges as one of the many one time (1T)
holographic pictures of the 4+2 dimensional string obtained via gauge fixing
of the 2T gauge symmetries. In 2T-physics the twistor language can be
transformed to usual spacetime language and vice-versa, off shell, as
different gauge fixings of the same 2T string theory. Further holographic
string pictures in 3+1 dimensions that are dual theories can also be
derived. The 2T superstring is further generalized in the SU(4)=SO(6) sector
of SU$\left( 2,2|4\right) $ by the addition of six bosonic dimensions, for a
total of 10+2 dimensions. Excitations of the extra bosons produce a SU$%
\left( 2,2|4\right) $ current algebra spectrum that matches the
classification of the high spin currents of $N=4,$ $d=4$ super Yang Mills
theory which are conserved in the weak coupling limit. This spectrum is
interpreted as the extension of the SU$\left( 2,2|4\right) $ classification
of the Kaluza-Klein towers of typeII-B supergravity compactified on AdS$%
_{5}\times $S$^{5}$, into the full string theory, and is speculated to have
a covariant 10+2 origin in F-theory or S-theory. Further generalizations of
the superstring theory to 3+2, 5+2 and 6+2 dimensions based on the
supergroups OSp$\left( 8|4\right) ,$ F$\left( 4\right) $, OSp$\left( 8^{\ast
}|4\right) $ respectively, and other cases, are also discussed. The OSp$%
\left( 8^{\ast }|4\right) $ case in 6+2 dimensions can be gauge fixed to 5+1
dimensions to provide a formulation of the special superconformal theory in
six dimensions either in terms of ordinary spacetime or in terms of twistors.

\section{2T formulation}

The Witten \cite{witten,witten2} and Berkovits \cite{berko,berko2} twistor
superstring, or the corresponding $N=4,$ $d=4$ super Yang-Mills, and
superconformal gravity theories \cite{berkWit}, are invariant under SO$(4,2)$
conformal symmetry and its supersymmetric generalization SU$\left(
2,2|4\right) $. The conformal symmetry SO$\left( d,2\right) $ is a hint for
Two-Time Physics (2T physics) \cite{2treviews} in \textit{flat} $d+2$
dimensional spacetime. In this paper it will be shown that the twistor
superstring is a gauge fixed version of a 2T superstring in 4+2 dimensions.

The first aim of the present paper is to apply to the Witten-Berkovits
twistor superstring the consequence of 2T-physics, which is the discovery of
dually related string models and the establishment of duality type relations
among them, while displaying an underlying spacetime with one extra space
and one extra time dimensions that unify the various dually related
holographic pictures as a single parent 2T theory.

Once this fact is established, the second aim of this paper is to propose
generalizations of the theory to $d+2$ dimensions. This works only in
special dimensions. The cases of $3+2,$ $4+2,$ $5+2,$ $6+2$ are direct
generalizations of the $4+2$ case and like the $4+2$ case also describe SYM
theory in dimensions $d=3,4,5,6$ respectively. Another type of
generalization of $4+2$ to $10+2$ is obtained by the addition of six more
dimensions to obtain a new 2T string theory in $10+2$ dimensions. In a
particular gauge of the 2T theory the $10+2$ system reduces to a $9+1$
dimensional theory and describes a string in the AdS$_{5}\times $S$^{5}$
background. The twistor version of this theory is also obtained. It is
already known that the particle limit of the $10+2$ string theory gives the
complete set of the Kaluza-Klein towers of type IIB supergravity
compactified on AdS$_{5}\times $S$^{5}$ as shown in \cite{2tAdSs}\cite%
{2tZero}. This is now generalized to the string version.

Although not yet fully analyzed, tentatively it appears that the full string
spectrum of the 10+2 theory has the SU$\left( 2,2|4\right) $ quantum numbers
that match the AdS-CFT correspondence to the conserved currents of $N=4,$ $%
d=4$ super Yang-Mills theory (SYM) in the weak coupling limit. The
classification of the $10+2$ states under the little group SO$\left(
10\right) $ was suggested a long time ago for the T-dual M-theory version in
$10+1$ dimensions \cite{11Dstring}, and is argued below that the SO$\left(
10\right) $ classification also applies to a 10+2 theory via T-duality. The
classification of the high-spin SYM currents at weak coupling was suggested
recently in \cite{bianchi} by using group theoretical steps that are closely
parallel to those previously used in \cite{11Dstring}\footnote{%
Note also that Polya theory as used in \cite{bianchi} is parallel to the
concept of ``color" as used in the oscillation representation of groups \cite%
{barsgunaydin}\cite{2tZero}.}. Most significantly, the SYM spectrum of
currents in \cite{bianchi} can be understood as the decomposition of SO$%
\left( 10\right) \rightarrow $SO$\left( 4\right) \times $SO$\left( 6\right) $
applied on the higher dimensional spectrum in \cite{11Dstring} (including
Kaluza-Klein excitations) corresponding to the compactification $\left(
10+2\right) \rightarrow \left( 4+2\right) +\left( 6+0\right) $ dimensions.
The 2T superstring in $10+2$ dimensions in the current paper seems to
provide the basis to explain this spectrum as belonging to a compactified
version of F-theory \cite{Ftheory} or S-theory \cite{Stheory}\cite{liftM},
thus giving the source of this spectrum in string theory. Furthermore, the
2T superstring taken in a variety of 1T gauges yields a collection of 1T
dynamical models as dual holographic pictures in 9+1 dimensions, all of
which have spectra related by dualities (changes of bases) within the same
group theoretical representations.

The concepts discussed in this paper are based on technology developed
previously in supersymmetric 2T-physics for particles \cite{2tSuper}, the
twistor gauge for 2T superparticles \cite{2ttwistor}, the 10+2 dimensional SU%
$\left( 2,2|4\right) $ superparticle \cite{2tAdSs}, the oscillator
representations of supergroups \cite{barsgunaydin} as refined recently \cite%
{2tZero}, and strings in the 2T framework \cite{2tString}. The salient
aspects of the previous work will be reviewed below.

A general remark about 2T-physics is that it contains a gauge symmetry Sp$%
(2,R)$ that, in flat spacetime{\footnote{%
The general case in the presence of backgrounds is developed in \cite%
{2tbackground}}}, acts on phase space $\left( X^{M},P^{M}\right) $ as a
doublet for each $M$. The first class constraints $X^{2}=P^{2}=X\cdot P=0$
due to this gauge symmetry have nontrivial solutions, and the theory is
unitary and causal, only if target spacetime has two timelike dimensions, no
more and no less. Then one finds that 2T-physics is consistent with one time
physics (1T-physics) in the Sp$\left( 2,R\right) $ gauge invariant sector,
or after the removal of gauge degrees of freedom. Thus, the two times in
target space arise as a consequence of the Sp$(2,R)$ gauge symmetry, they
are not an input. Then SO$(d,2)$ is the global symmetry acting on the
spacetime $M$ index in flat space. The SO$(d,2)$ symmetry commutes with the
local Sp$(2,R),$ hence it is gauge invariant, and classifies the physical
spectrum.

Another general remark about 2T-physics is that many one-time physics
(1T-physics) systems emerge via gauge fixing the 2T system. The 2T action
naturally unifies 1T systems that are dually related among themselves,
because they are all related to the same parent 2T system via the Sp$(2,R)$
gauge transformations. Thus 1T systems, which may appear unrelated in the
absence of the understanding reached via 2T-physics, get unified through the
higher dimensional 2T theory. The gauge fixing from 2T to 1T is done by
using two out of the three local parameters of the Sp$\left( 2,R\right) $
symmetry, plus two out of the three corresponding constraints, to reduce the
phase space degrees of freedom by one timelike and one spacelike degrees of
freedom. The remaining phase space has $\left( d-1\right) $ spacelike and $1$
timelike dimensions, and provides a holographic description of the higher
dimensional 2T system in the reduced phase space. There are many possible
holographic pictures of the higher system corresponding on how the remaining
timelike coordinate is embedded in the $d+2$ higher spacetime.

The choice of the remaining 1-timelike coordinate rearranges the dynamics of
the 2T system to evolve according to that choice of time. This leads to
different Hamiltonians to describe each of the 1T holographic pictures.
Therefore, each distinct gauge choice of time makes the same 2T system
appear as distinct dynamics from the point of view of 1T physics. The
holographic pictures obtained from the same 2T action are related to one
another by duality type relations, where the duality transformation is an Sp$%
\left( 2,R\right) $ gauge transformation (which is non-linearly realized on
the remaining phase space once a gauge is chosen). Many striking examples of
this phenomenon have been displayed in simple classical and quantum
mechanics. Some of the simplest examples include free relativistic particle,
its twistor description, non-relativistic particle, hydrogen atom, harmonic
oscillator, particle on AdS$_{d-k}\times $S$^{k}$ background, etc., all
being holographic pictures of the same 2T \textit{free particle}. One can
directly verify that these systems are indeed related as predicted by
2T-physics \cite{2tHandAdS}\cite{2treviews}. These simple examples, and
spinning and supersymmetric generalizations, with or without background
fields \cite{2treviews,2tbackground}, already establish the existence of
2T-physics as a solid framework that describes reality.

The global symmetry SO$(d,2)$ in flat spacetime is linearly realized in the
2T phase space $\left( X^{M},P^{M}\right) $. In the present paper $M$ is
labelled as $M=\left( 0^{\prime },0,1^{\prime },1,2,\cdots ,\left(
d-1\right) \right) ,$ or $M=\left( +^{\prime },-^{\prime },\mu \right) $
where $+^{\prime },-^{\prime }$ are lightcone type combinations constructed
from the directions $0^{\prime },1^{\prime },$ while $\mu =\left( 0,1,\cdots
,\left( d-1\right) \right) $ is a $d$-dimensional Lorentz index.

\section{2T superparticle in 4+2}

The 2T formulation of the superparticle in $d=3,4,5,6$ with $N$
supersymmetries is introduced in \cite{2tSuper} and further developed in
\cite{2ttwistor}, while the extension with more bosonic dimensions is
discussed in \cite{2tAdSs}. These will be directly relevant for the 2T
reformulation of the twistor superstring. For this purpose, first we recall
the superparticle in $d=4$ dimensions with $N=4$ supersymmetries. It
requires $4+2$ coordinates $X^{M}\left( \tau \right) $ and momenta $%
P^{M}\left( \tau \right) ,$ and a supergroup element $g\left( \tau \right)
\in $SU$\left( 2,2|4\right) $ that contains fermions $\Theta _{s}^{a}\left(
\tau \right) $ in the off-diagonal blocks, where $\left( s,a\right) $ denote
the complex bi-fundamental representation $\left( 4,4\right) $ of SU$\left(
2,2\right) \times $SU$\left( 4\right) .$ This spinor has double the size of
the smallest spinor $\theta _{\alpha }^{a}\left( \tau \right) $ in $4$
dimensions, which is of course necessary if the SO$\left( 4,2\right) =$SU$%
\left( 2,2\right) $ is to be realized linearly in 4+2 dimensions. Thus,
compared to the 1T formulation there are extra degrees of freedom in $%
X,P,\Theta $ and in the SU$\left( 2,2\right) \times $SU$\left( 4\right) $
bosonic blocks in $g\left( \tau \right) .$ If the covariant 2T formulation
of the superparticle is to be equivalent to the 1T formulation there has to
be various gauge symmetries and extended kappa supersymmetries to cut down
the degrees of freedom to the correct set for the superparticle in $d=4$ and
$N=4$. As shown in \cite{2tSuper}\cite{2ttwistor} this is beautifully
achieved with the following action
\begin{equation}
\mathcal{L}_{2T}=\frac{1}{2}\varepsilon ^{ij}\partial _{\tau }X_{i}\cdot
X_{j}-\frac{1}{2}A^{ij}X_{i}\cdot X_{j}+\frac{1}{2}Str\left( i\partial
_{\tau }gg^{-1}L\right) ,\;\;L\equiv \left(
\begin{array}{cc}
\frac{i}{2}\Gamma _{MN}L^{MN} & 0 \\
0 & 0%
\end{array}%
\right) ,  \label{L}
\end{equation}%
where $X_{i}^{M}=\left( X^{M},P^{M}\right) ,$ with $i=1,2,$ is the Sp$\left(
2,R\right) $ doublet; $\varepsilon ^{ij}$ is the antisymmetric invariant
metric of Sp$\left( 2,R\right) $; the symmetric $A^{ij}=A^{ji}$ is the Sp$%
\left( 2,R\right) $ gauge potential; $\Gamma _{M}$ and $\Gamma _{MN}=\frac{1%
}{2}\left[ \Gamma _{M},\Gamma _{N} \right] $ are SO$\left( 4,2\right) $
gamma matrices in the spinor representation embedded in the first $4\times 4$
SU$\left( 2,2\right) $ block of the SU$\left( 2,2|4\right) $ matrix. The
Cartan connection $i\partial _{\tau }gg^{-1}$ projected in the direction of
the subgroup SO$\left( 4,2\right) \in $SU$\left( 2,2|4\right) $ is coupled
to the SO$\left( 4,2\right) $ orbital angular momentum $L^{MN}=\varepsilon
^{ij}X_{i}^{M}X_{j}^{N}=X^{M}P^{N}-X^{N}P^{M}$ which is Sp$\left( 2,R\right)
$ gauge invariant. Note that this SO$\left( 4,2\right) $ connection is not a
pure gauge since the 16 complex fermionic coset parameters $\Theta _{s}^{a}$
in SU$\left( 2,2|4\right) /($SU$\left( 2,2\right) \times $SU$\left( 4\right)
)$ contribute to it.

This action has a \textit{global} SU$\left( 2,2|4\right) _{R}$ supersymmetry
that acts on the right side of $g,$ namely $g\left( \tau \right) \rightarrow
g\left( \tau \right) g_{R}.$ It also has \textit{local} bosonic SU$\left(
2,2\right) \times $SU$\left( 4\right) $ and local fermionic kappa
supersymmetries embedded in SU$\left( 2,2|4\right) _{L}$ that act
simultaneously on the left side\footnote{$g_{L}\left( \tau \right) $ has
some restrictions on the fermionic parameters as explained in more detail in
\cite{2ttwistor} and the section on the string below. Therefore $g_{L}\left(
\tau \right) $ cannot remove all of the degrees of freedom from $g\left(
\tau \right) $ since it contains fewer independent fermionic parameters than
SU$\left( 2,2|4\right) _{L}$.} of $g,$ namely $g\left( \tau \right)
\rightarrow g_{L}\left( \tau \right) g\left( \tau \right) $, as well as on
the phase space $\left( X^{M},P^{M}\right) $ and the gauge fields $A^{ij}$
(see Eqs.(\ref{gR}-\ref{34})).

We discuss below three approaches to the quantization of this 2T
superparticle system: covariant quantization in 4+2 dimensions, 1T
superparticle gauge in 3+1 dimensions and its quantization, and the
supertwistor gauge and its quantization. In every approach the physical
quantum states correspond to the physical degrees of freedom of $N=4$, $d=4$
SYM theory.

The conserved Noether charges for the global SU$\left( 2,2|4\right) _{R}$
symmetry are the entries of the $8\times 8$ supermatrix\footnote{%
The definition of $L$ and $J$ in this paper differ from the ones in \cite%
{2tZero}\cite{2tAdSs} by overall factors. Consequently the formulas
involving $J$ in the present paper differ by the corresponding modifications
from those of \cite{2tZero}\cite{2tAdSs}.}
\begin{equation}
J_{A}^{~B}=\left( \frac{1}{4}g^{-1}Lg\right) _{A}^{~B},\;\;\partial _{\tau
}J_{A}^{~B}\left( \tau \right) =0.  \label{J}
\end{equation}%
These charges are invariant under all of the gauge symmetries, namely Sp$%
\left( 2,R\right) $ as well as the ones embedded in SU$\left( 2,2|4\right)
_{L}$ (more details on this when we discuss the string below). Therefore
they define the physical states of the system as representations of SU$%
\left( 2,2|4\right) _{R}.$ Let us analyze some properties of $J.$ The square
of this matrix is $J^{2}=\frac{1}{16}g^{-1}L^{2}g.$ At the classical level $%
L^{2}$ vanishes after using the Sp$\left( 2,R\right) $ constraints $%
X_{i}\cdot X_{j}=0$ that follow from the action (namely when one considers
the Sp$\left( 2,R\right) $ gauge invariant sector of phase space $%
X^{M},P^{M} $). At the quantum level, we use the commutation rules of $%
L^{MN} $ to compute $L^{2}=\left( D-2\right) L+diag\left( \frac{1}{2}%
L^{MN}L_{MN}~\mathbf{1}_{2,2}~,\mathbf{0}_{4}\right) ,$ including the linear
term. Thus, at the quantum level, instead of vanishing $J^{2}$ we obtain for
$D=6$ the following projector equation
\begin{equation}
\left( J^{2}\right) _{A}^{~B}=J_{A}^{~B}
\end{equation}%
on the Sp$\left( 2,R\right) $ gauge invariant quantum states\footnote{%
It must be mentioned that, due to constraints, $\frac{1}{2}L^{MN}L_{MN}$ may
not commute with $g$ at the quantum level. Since this factor is sandwiched
between $g^{-1}$ and $g,$ it must be passed through $g$ before it is applied
on Sp$\left( 2,R\right) $ invariant physical states. Only after this step,
for $g\in $SU$\left( 2,2|4\right) ,$ one finds that the term involving $%
\frac{1}{2}L^{MN}L_{MN}$ vanishes on Sp$\left( 2,R\right) $ gauge invariant
physical states. This result is verified by quantizing the system in fixed
gauges, as seen below easily in the twistor gauge.}. This gives the SU$%
\left( 2,2|4\right) _{R}$ covariant quantization of Eq.(\ref{L}), by
identifying the physical states as those that satisfy the condition $J^{2}=J$
for the SU$\left( 2,2|4\right) _{R}$ charges. The supersingleton of SU$%
\left( 2,2|4\right) _{R}$ satisfies this condition (this was first realized
in the context of 2T physics in \cite{2tZero}\cite{2tAdSs}). Furthermore, it
is well known that the spectrum of the SU$\left( 2,2|4\right) $
supersingleton corresponds precisely to the fields of the linearized $N=4$, $%
d=4$ super Yang-Mills theory (SYM) and all their derivatives \cite%
{barsgunaydin}\cite{2tZero}. Hence the physical states of Eq.(\ref{L}) are
described by the SYM fields.

It is also possible to obtain the supersingleton spectrum by choosing some
gauge which gives a holographic picture in 3+1 dimensions. The action Eq.(%
\ref{L}) has just the required gauge symmetries to gauge fix the 2T
superparticle into various holographic pictures that describe 1T-physics.
One holographic picture is the standard superparticle in 4 dimensions
\begin{equation}
\mathcal{L}_{particle}=\dot{x}\cdot p-\frac{1}{2}A^{22}p^{2}+\tilde{\theta}%
_{a}\gamma ^{\mu }\partial _{\tau }\theta ^{a}p_{\mu }\leftrightarrow \frac{1%
}{2A^{22}}\left( \dot{x}^{\mu }+\bar{\theta}_{a}\gamma ^{\mu }\partial
_{\tau }\theta ^{a}\right) ^{2}.  \label{L1}
\end{equation}%
with $\mu =0,1,2,3,$ and $\theta _{\alpha }^{a}$ four complex SL$\left(
2,C\right) $ doublets. This holographic picture is generated by (1) using a
local symmetry SU$\left( 2,2\right) \times $SU$\left( 4\right) \subset $SU$%
\left( 2,2|4\right) _{L}$ to remove all the bosonic degrees of freedom in $%
g\left( \tau \right) ;$ (2) partially fixing the fermionic kappa symmetry in
SU$\left( 2,2|4\right) _{L}$ to cut down the original 16 complex fermions $%
\Theta _{s}^{a}$ in $g$ by a factor of two, to 8 complex fermions $\theta
_{\alpha }^{a}$, i.e. $\alpha=1,2; \ a=1,2,3,4$ (with leftover kappa
symmetry), so that $g\left( \tau \right) $ takes the form%
\begin{equation}
g=\exp \left(
\begin{array}{ccc}
0_{2} & 0 & \theta \\
0 & 0_{2} & 0 \\
0 & \bar{\theta} & 0_{4}%
\end{array}%
\right) =\left(
\begin{array}{ccc}
1_{2} & \frac{1}{2}\theta \bar{\theta} & \theta \\
0 & 1_{2} & 0 \\
0 & \bar{\theta} & 1_{4}%
\end{array}%
\right) ;  \label{gfixed}
\end{equation}%
(3) partially fixing the Sp$\left( 2,R\right) $ gauge symmetry by choosing
the $M=+^{\prime }$ doublet in the form $\left( X^{+^{\prime
}}=1,P^{+^{\prime }}=0\right) ,$ and solving two of the constraints $%
X^{2}=X\cdot P=0$ to reduce the 2T phase space $\left( X^{M},P^{M}\right) $
to the gauge that describes the relativistic particle $\left( x^{\mu
},p^{\mu }\right) $ in $d=4$, namely $X^{+^{\prime }}=1$, $X^{-^{\prime
}}=x^{2}/2$, $X^{\mu }=x_{\mu }$ and $P^{+^{\prime }}=0$, $P^{-^{\prime
}}=x.p$, $P^{\mu }=p^{\mu }.$ Then the 2T system in Eq.(\ref{L}) reduces to
the 1T superparticle in Eq.(\ref{L1}) \cite{2tSuper}\cite{2ttwistor}.

As is well known, the on-shell quantum states of the superparticle described
by Eq.(\ref{L1}) is given by the on-shell fields of linearized $N=4$ SYM. A
quick way of understanding this is by performing quantization in the
lightcone gauge, which gives $2^{3}$ bosons and 2$^{3}$ fermions\footnote{%
After gauge fixing the remaining kappa supersymmetry, 8 real components of
the $\theta $'s remain. Upon quantization, they satisfy a Clifford algebra
which is realized on $2^{4}$ states, i.e. $2^{3}$ bosons plus $2^{3}$
fermions.}, with on shell momenta in 3+1 dimensions $|2_{B}^{3},p>\oplus
|2_{F}^{3},p>$. These $8_{Bose}+8_{Fermi}$ quantum states of the
superparticle correspond to the transverse physical fields of $N=4,$ $d=4$
SYM in the lightcone gauge with helicities (in parentheses) times their SU$%
\left( 4\right) $ multiplicities given by
\begin{equation}
\left( +1\right) \oplus \left( +1/2\right) \times 4\oplus \left( 0\right)
\times 6\oplus \left( -1/2\right) \times \bar{4}\oplus \left( -1\right) .
\end{equation}%
Thus the $8_{Bose}+8_{Fermi}$ states, taken in position space, correspond to
the on-shell SYM fields in the lightcone gauge that are classified by SO$%
\left( 2\right) \subset $SO$\left( 3,1\right) $ as the little group that
describes the helicities: $A_{i}\left( x\right) $ with $i=1,2$ for the SO$%
\left( 2\right) $ vector in transverse directions, $\psi _{1/2}^{a}\left(
x\right) ,\bar{\psi}_{-1/2,a}\left( x\right) $ for the SO$\left( 2\right) $
fermions in the $4$ and $\bar{4}$ representations of SU$\left( 4\right) ,$
and $\phi ^{\lbrack ab]}\left( x\right) $ for the SO$\left( 2\right) $
scalars in the 6 dimensional antisymmetric tensor of SU$\left( 4\right) .$
In this holographic picture, the original SU$\left( 2,2|4\right) _{R}$
global supersymmetry in Eq.(\ref{L}) becomes the non-linearly realized $N=4$
superconformal symmetry, both of the gauge fixed action in Eq.(\ref{L1}) and
of the $N=4$ SYM action.

In 2T-physics each gauge may appear to describe various 1T-physics systems
as holographic pictures in 3+1 dimensions, but the representation of the
gauge invariant SU$\left( 2,2|4\right) _{R}$ cannot change by choosing some
gauge since $J$ is gauge invariant. Hence, it must be true that the rather
complicated non-linear representation of the superconformal supergroup SU$%
\left( 2,2|4\right) _{R}$ for the superparticle \cite{schwarz}\cite{2tSuper}%
, properly operator ordered at the quantum level, must satisfy the projector
condition $J^{2}=J.$ This is guaranteed by its 2T-physics origin in the
gauge invariant form $J=\frac{1}{4}g^{-1}Lg,$ which is then gauge fixed by
inserting the gauge fixed versions of $g,X,P$ given above \cite{2ttwistor}.

Another gauge fixed form is the twistor description of the superparticle as
discussed in \cite{2ttwistor}. This is done by using the Sp$\left(
2,R\right) $ symmetry and the SU$\left( 2,2\right) \subset $SU$\left(
2,2|4\right) _{L}$ (that also locally rotates phase space $X_{i}^{M}$ as SO$%
\left( 4,2\right) ;$ see string case below) to completely fix $X^{M},P^{M}$
to the form $X^{+^{\prime }}=1$ and $P^{-}=1$ (note, not $P^{-^{\prime }}$)
while all other components vanish. These $X^{M},P^{M}$ already satisfy the
constraints $X^{2}=P^{2}=X\cdot P=0$. In this gauge the only non-vanishing
component of $L^{MN}$ is $L^{+^{\prime }-}=1$. Hence the 2T action in Eq.(%
\ref{L}) and the SU$\left( 2,2|4\right) _{R}$ charges in Eq.(\ref{J}) become%
\begin{eqnarray}
\mathcal{L}_{twistor} &=&-\frac{1}{4}Str\left( \partial _{\tau
}gg^{-1}\Gamma \right) =\bar{Z}^{A}\left( \partial _{\tau }Z_{A}\right) ,
\label{Ltwistor} \\
\left( J\right) _{A}^{B} &=&\left( \frac{1}{4}g^{-1}\Gamma g\right)
_{A}^{B}=Z_{A}\bar{Z}^{B}.\;\;\Gamma \equiv \left(
\begin{array}{cc}
\Gamma _{+^{\prime }-} & 0 \\
0 & 0%
\end{array}%
\right)  \label{JZZ}
\end{eqnarray}%
These twistor forms arise from one row of the matrix $g$ and one column of
the matrix $g^{-1}$ since $\Gamma$ has only one nonzero off-diagonal entry.
It is evident from Eq.(\ref{Ltwistor}) that $Z_{A},\bar{Z}^{A}$ are
canonically conjugate complex supertwistors which can be expressed in terms
of oscillators\footnote{%
We use the notation of \cite{2tZero} to identify the oscillators in the SU$%
\left( 2\right) \times $SU$\left( 2\right) \times $SU$\left( 4\right) $
\textit{unitary} basis of SU$\left( 2,2|4\right).$ They are $Z_{A}=\left(
\begin{array}{c}
a_{n} \\
\bar{b}^{m} \\
\psi _{r}%
\end{array}%
\right) ,\;\;\bar{Z}^{A}=\left( \bar{a}^{n},-b_{m},\bar{\psi}^{r}\right) ,$
where a bar (such as $\bar{a}^{n}$) means creation operator, and otherwise
annihilation operator. The extra minus sign in $\bar{Z}^{A}$ is due to the SU%
$\left( 2,2\right) $ metric in the SU$\left( 2\right) \times $SU$\left(
2\right) $ basis, and it is the reason for having an annihilation operator $%
b $ for that entry instead of a creation operator (the canonical structure
is imposed by the corresponding signs in the action). The indices take the
values $n=1,2,~m=1,2,$ $r=1,2,3,4$. Then, after reordering the oscillators $%
\bar{Z}^{A}Z_{A}=\bar{a}\cdot a-\left( \bar{b}\cdot b+2\right) +\bar{\psi}%
\cdot \psi ,$ we write the constraint $\bar{Z}^{A}Z_{A}=0$ in terms of the
number operators in the form $\Delta \equiv N_{a}-N_{b}+N_{\psi }=2.$ This
is the $\Delta =2$ condition (for one color) in \cite{2tZero} that gave the
SYM states as the supersingleton.\label{oscillators}}. Due to their
embedding in the supergroup element $g,$ the supertwistors must satisfy $%
\bar{Z}^{A}Z_{A}=0,$ a condition which arises from an off diagonal entry in $%
gg^{-1}=1$. Furthermore, the condition $\bar{Z}^{A}Z_{A}=0$ corresponds to
Str$\left( J\right) =Z_{A}\bar{Z}^{A}\left( -1\right) ^{A}=\bar{Z}%
^{A}Z_{A}=0 $. The constraint $\bar{Z}^{A}Z_{A}=0$ may also be interpreted
as arising from a gauge symmetry U$\left( 1\right) $ of Eq.(\ref{L}) as part
of the original gauge symmetries in SU$\left( 2,2|4\right) _{L}.$ Note the
change of orders of operators in $Z_{A}\bar{Z}^{A}\left( -1\right) ^{A}=\bar{%
Z}^{A}Z_{A}$ is valid at the quantum level without any constant residues in
the case of SU$\left( n,n|2n\right) $.

The quantum states generated by the supertwistors are precisely the ones
described by the well known oscillator representation of the supergroup SU$%
\left( 2,2|4\right) _{R}$ \cite{barsgunaydin}\cite{2tZero} with the
additional condition $\bar{Z}^{A}Z_{A}=0$ that selects the physical states
in super Fock space. These oscillator states correspond precisely to the
supersingleton representation of SU$\left( 2,2|4\right) _{R}$ that describes
N=4 SYM. Again, the charges $J$ in this gauge satisfy the projector
condition $J^{2}=J$ on physical states. This is easily verified directly in
this gauge by using Eq.(\ref{JZZ})
\begin{equation}
\left( J^{2}\right) _{A}^{B}=Z_{A}\bar{Z}^{C}Z_{C}\bar{Z}^{B}=Z_{A}\bar{Z}%
^{B}\left( \bar{Z}^{C}Z_{C}+1\right) \underset{phys.~states}{=}Z_{A}\bar{Z}%
^{B}=\left( J\right) _{A}^{B}  \label{projector}
\end{equation}%
In this computation we have used the oscillator commutation rules to pass
the number operator $\bar{Z}^{C}Z_{C}$ through $\bar{Z}_{B}$ and then set $%
\bar{Z}^{C}Z_{C}=0$ on physical states (kets), thus showing that the
projector condition $J^{2}=J$ is true on physical states.

One may choose other holographic pictures of the same 2T system, with
varying physical interpretations of the 1T systems that arise in various
gauges. For example, the $N=4$ AdS$_{2}\times $S$^{2}$ superparticle (SO$%
\left( 1,2\right) \times $SO$\left( 3\right) $ basis) and the $N=4$ AdS$%
_{3}\times $S$^{1}$ superparticle (SO$\left( 2,2\right) \times $SO$\left(
2\right) $ basis) emerge as duals to the supersymmetric particle (SO$\left(
1,1\right) \times $SO$\left( 3,1\right) $ basis) or the supertwistor system
given above. Other examples of interest are the $N=4$ supersymmetric
Hydrogen atom in three space dimensions (SO$\left( 2\right) \times $SO$%
\left( 4\right) $ or SO$\left( 1,2\right) \times $SO$\left( 3\right) $
bases), and the $N=4$ harmonic oscillator in two space dimensions (SO$\left(
2,2\right) \times $SO$\left( 2\right) $ basis), which also emerge from gauge
choices of Eq.(\ref{L}). The purely bosonic versions of these examples (and
some other generalizations) are discussed in detail in \cite{2tHandAdS}\cite%
{2treviews}\footnote{%
Another recent application of the 2T-physics approach is the formulation of
the adjoint representation of the high spin algebra in terms of phase space $%
\left( X^{M},P^{M}\right) $ in any dimension \cite{vasiliev}. As it should
be expected, it corresponds to the SO$\left( d,2\right) $ singleton, whose
quadratic Casimir $C_{2}=1-d^{2}/4$ in any dimension was computed in \cite%
{2treviews}. The SO(2)$\times $SO$\left( d\right) $ basis described in \cite%
{vasiliev} in terms of oscillators is an equivalent description of the phase
space SO(2)$\times $SO$\left( d\right) $ basis of the H-atom gauge given in
\cite{2tHandAdS}. Applying the same methods, the work of \cite{vasiliev} is
generalized to the supersymmetric version of the high spin algebra hs$%
(2,2|4) $ through our SU$\left( 2,2|4\right) $ 2T system of Eqs.(\ref{L},\ref%
{J}), either covariantly, or taken in a variety of gauges, all of which
describe the supersingleton. The supersymmetric generalization for high spin
can be done also for the other dimensions discussed in this paper.} at the
classical and quantum levels. Each one of these $N=4$ systems is represented
by the SU$\left( 2,2|4\right) _{R}$ supersingleton rearranged in various
bases; hence each has a spectrum that is dual to the $N=4$ SYM spectrum. The
SU$\left( 2,2|4\right) $ symmetry is interpreted as conformal symmetry in
the superparticle gauge (SO$\left( 1,1\right) \times $SO$\left( 3,1\right) $
basis), but it has other interpretations as a nonlinear hidden symmetry in
the other cases.

The interacting $N=4,d=4$ SYM theory, rewritten in the appropriate basis,
may be taken as an interesting interacting theory for any of the 1T
holographic pictures. There should also be a field theoretic formulation of
this theory directly written covariantly in 4+2 dimensions. The projector
equation $J^2=J$ is very suggestive as an equation of motion of cubic string
field theory, and one may develop an interacting field theory along those
lines for the 2T superparticle after introducing ghosts and a BRST operator {%
\footnote{%
See also the 2T-physics field theory approaches along along the lines of
\cite{2tfield}.}. }

\section{2T superstring in 4+2\label{4+2}}

We now present an action for a superstring in 2T-physics in $4+2$
dimensions. This action has many holographic pictures in $3+1$ dimensions,
with varying 1T physical interpretations, that parallel those of the 2T
superparticle of the previous section. One of them is the twistor
superstring.

The worldsheet `` matter"\ fields are $X^{M}\left( \tau ,\sigma \right)
,\left( P^{m}\left( \tau ,\sigma \right) \right) ^{M}$, and the SU$\left(
2,2|4\right) $ supergroup element $g\left( \tau ,\sigma \right) ,$ which are
the string analogs of the particle case, while the analogs of the three Sp$%
\left( 2,R\right) $ gauge fields $A^{ij}=\left( A^{11},A^{22},A^{12}\right) $
are now $\left( A\left( \tau ,\sigma \right) ,B_{mn}\left( \tau ,\sigma
\right) ,C_{m}\left( \tau ,\sigma \right) \right) $ respectively. The action
is

\begin{equation}
\sqrt{-\gamma }\mathcal{L}^{-}=\partial _{m}X\cdot P^{-m}-\frac{1}{2}AX\cdot
X-\frac{1}{2}B_{mn}P^{-m}\cdot P^{-n}-C_{m}P^{-m}\cdot X+\frac{1}{2}%
Str\left( i\partial _{m}gg^{-1}L^{-m}\right) +\mathcal{L}_{1}^{-},
\label{Ls}
\end{equation}%
with
\begin{equation}
L^{-m}\equiv \left(
\begin{array}{cc}
\frac{i}{2}\Gamma ^{MN}X_{[M}P_{N]}^{-m} & 0 \\
0 & 0%
\end{array}%
\right) .  \label{Lm}
\end{equation}%
Here $\left( P^{-m}\right) ^{M}$ is the chirally projected component of the
worldsheet momentum current density $\left( P^{-m}\right) ^{M}=\frac{1}{2}%
\left( \sqrt{-\gamma }\gamma ^{mn}-\varepsilon ^{mn}\right) P_{m}^{M},\;$%
where $\gamma _{mn}$ is the worldsheet metric and $\varepsilon ^{mn}$ is the
constant antisymmetric tensor. In what follows, it is important to realize
that the projected $P^{-m}$ has only one independent component on the
worldsheet, since the opposite projector $\left( \sqrt{-\gamma }\gamma
^{mn}+\varepsilon ^{mn}\right) /2$ annihilates it. $\mathcal{L}%
_{1}^{-}\equiv \mathcal{L}_{1}^{-}\left( A,B,C,\gamma ,j^{r}\right) $ is an
additional part of the action that contains the current $j_{m}^{r}\left(
\tau ,\sigma \right) $ of the twistor superstring \cite{berko,berko2}, and
perhaps the other fields. It will not be necessary to discuss details of $%
\mathcal{L}_{1}^{-}$ in this paper. Note that factors of $\sqrt{-\gamma }$
are already absorbed into the definition of the gauge fields $\left(
A,B_{mn},C_{m}\right) .$

One may also introduce a Lagrangian $\mathcal{L}^{+}$ with the opposite
worldsheet chirality projectors obtained from $\mathcal{L}^{-}$ by replacing
$\varepsilon ^{mn}\rightarrow -\varepsilon ^{mn}$. It appears that one may
formulate the twistor superstring either as an open string with both $%
\mathcal{L}^{+}+\mathcal{L}^{-}$ and open string boundary conditions, or as
a closed string with only $\mathcal{L}^{-}$ \cite{siegel}. We will
concentrate on the latter approach and hence study the properties of $%
\mathcal{L}^{-}$ in the rest of this paper\footnote{%
A related purely bosonic string action, without the projectors $\left( \sqrt{%
-\gamma }\gamma ^{mn} \pm \varepsilon ^{mn}\right)/2$, was considered in
\cite{2tString} in $d+2$ dimensions. The conclusion in \cite{2tString} was
that the solution space of the 2T string reduced to a tensionless string in $%
d$ dimensions after gauge fixing. Although the notation was slightly
different, an action equivalent to the one in \cite{2tString} is written in
the present notation by dropping the projectors so that all components of $%
P^{m}$ and all components of the gauge fields ($A,B_{mn},C_{m}$) contribute.
It turns out that the extra equation of motion $P^{+}\cdot P^{-}=0$ that
would follow from varying $B_{+-}$ is the one responsible for imposing
tensionless strings for the solution space. With only $\mathcal{L}^{-}$ or $%
\mathcal{L}^{+}+\mathcal{L}^{-}$ this condition is avoided since $B_{+-}$ is
absent. In any case, tensionless strings play a role in the overall scheme.}.

The action $S=\int d\tau d\sigma \sqrt{-\gamma }\mathcal{L}^{-}$ is clearly
invariant under reparametrizations of the worldsheet. In the conformal gauge
$\gamma _{mn}=\gamma \eta _{mn},$ it is convenient to choose coordinates $%
\sigma ^{\pm }=\tau \pm \sigma ,$ and basis $\eta _{+-}=1,~\eta _{\pm \pm
}=0,$ with $m,n=\pm $. The Lagrangian $\mathcal{L}^{-}$ contains only the $%
m=n=-$ components $P^{-},B_{--},C_{-},$ and takes the form
\begin{equation}
\sqrt{-\gamma }\mathcal{L}^{-}=\partial _{-}X\cdot P^{-}-\frac{1}{2}AX\cdot
X-\frac{1}{2}B_{--}P^{-}\cdot P^{-}-C_{-}P^{-}\cdot X+\frac{1}{2}Str\left(
i\partial _{-}gg^{-1}L^{-}\right) +\mathcal{L}_{1}^{-}.  \label{Lss}
\end{equation}%
This looks just like the particle counterpart in Eq.(\ref{L}) with $%
A^{11}\rightarrow A,$ $A^{22}\rightarrow B_{--}$ , $A^{12}\rightarrow C_{-}$
and $P\rightarrow P^{-},$ and therefore has the same structure of
symmetries, but now local on the worldsheet instead of the worldline. Hence
holographic pictures of the $4+2$ dimensional 2T string are obtained in $3+1$
dimensions by gauge choices just as in the particle case. One of these
holographic pictures is the Witten-Berkovits twistor superstring.

Let us first discuss the symmetries in more detail in the case of the string
in Eq.(\ref{Ls}) before any gauge choices. There are three kinds of
symmetries as itemized below.

\begin{itemize}
\item[1-] The SU$\left( 2,2|4\right) _{R}$ global symmetry of the particle
case is replaced by the transformation
\begin{equation}
g\rightarrow g^{\prime }=gg_{R},\;\partial _{-m}g_{R}=0,  \label{gR}
\end{equation}%
indicating that $g_{R}$ is not a constant, but is \textit{holomorphic}. The
fields $X,P^{m},A,B_{mn},C_{n},\gamma _{mn}$ are neutral under this
symmetry. The conserved Noether current for this symmetry is
\begin{equation}
J^{-m}=\frac{1}{4}g^{-1}L^{-m}g,\;\;\partial _{m}J^{-m}=0.  \label{JR}
\end{equation}%
The conservation is verified through the equations of motion. This current
corresponds to a SU$\left( 2,2|4\right) $ Kac-Moody algebra whose
representations classify the physical states of the theory.\qquad
\end{itemize}

The three components of the Sp$(2,R)$ transformations in the particle case
are replaced by the transformations $\delta _{\alpha },\delta _{\rho
},\delta _{\beta }$, as follows:

\begin{itemize}
\item[2a-] Local dilatations $\delta _{\alpha }$:
\begin{equation}
\delta _{\alpha }X=\alpha X,\;\delta _{\alpha }P^{m}=-\alpha P^{m},\;\delta
_{\alpha }A=-2A\alpha ,\;\delta _{\alpha }B_{mn}=2B_{mn}\alpha ,\;\delta
_{\alpha }C_{m}=\partial _{m}\alpha .  \label{21}
\end{equation}%
The $\gamma _{mn}$ and $g$ fields are neutral under $\delta _{\alpha }.$
Then we obtain $\delta _{\alpha }L^{-m}=0,~$ $\delta _{\alpha }J^{m}=0,$ and
$\delta _{\alpha }\left( \sqrt{-\gamma }\mathcal{L}^{-}\right) =0.$

\item[2b-] Local $\rho $-transformations $\delta _{\rho }$:
\begin{equation}
\delta _{\rho }X=0,\;\delta _{\rho }P^{m}=-\rho ^{m}X,\;\delta _{\rho
}A=2C_{m}\rho ^{m}+\partial _{m}\rho ^{m},\;\delta _{\rho }B_{mn}=0,\;\delta
_{\rho }C_{m}=B_{mn}\rho ^{n}.  \label{22}
\end{equation}%
The $\gamma _{mn}$ and $g$ fields are neutral under $\delta _{\rho }.$ Then $%
\delta _{\rho }L^{-m}\sim \Gamma ^{MN}X_{[M}X_{N]}\rho ^{-m}=0,$ $\delta
_{\rho }J^{-m}=0,$ while $\delta _{\rho }\left( \sqrt{-\gamma }\mathcal{L}%
^{-}\right) =-\frac{1}{2}\partial _{m}\left( \rho ^{-m}X\cdot X\right) $ is
a total derivative.

\item[2c-] Local $\beta $-transformations $\delta _{\beta }$:
\begin{equation}
\delta _{\beta }X=\beta _{m}P^{-m},\;\delta _{\beta }P^{m}=0,\;\delta
_{\beta }A=0,\;\delta _{\beta }B_{mn}=-C_{(m}\beta _{n)}+\frac{1}{2}\partial
_{(m}\beta _{n)},\;\delta _{\beta }C_{m}=-A\beta _{m}.  \label{23}
\end{equation}%
The $\gamma _{mn}$ and $g$ fields are neutral under $\delta _{\beta }.$ Then
$\delta _{\beta }L^{-m}\sim \Gamma ^{MN}P_{[M}^{-m}P_{N]}^{-n}\beta _{n}=0$
(since the projected index $\left( -m\right) $ can take only one value).
This gives again $\delta _{\rho }J^{-m}=0,$ while $\delta _{\beta }\left(
\sqrt{-\gamma }\mathcal{L}^{-}\right) =\partial _{m}\left( \frac{1}{2}\beta
_{n}P^{-m}\cdot P^{-n}\right) $ is a total derivative.
\end{itemize}

The local symmetries embedded in SU$\left( 2,2|4\right) _{L}$ are as follows:

\begin{itemize}
\item[3a-] There is a local SO$\left( 4,2\right) =$SU$\left( 2,2\right) $
Lorentz symmetry with parameters $\varepsilon ^{MN}\left( \tau ,\sigma
\right) $ under \textit{left multiplication} of $g$ in the spinor
representation and simultaneous transformation of $X^{M},\left(
P^{-m}\right) ^{M}$ in the vector representation
\begin{equation}
\delta _{\varepsilon }X^{M}=\varepsilon ^{MN}X_{N},\quad \delta
_{\varepsilon }\left( P^{-m}\right) ^{M}=\varepsilon ^{MN}\left(
P_{M}^{-m}\right) ,\;\delta _{\varepsilon }g=\frac{1}{4}\varepsilon
^{MN}\left(
\begin{array}{cc}
\Gamma ^{MN} & 0 \\
0 & 0%
\end{array}%
\right) \,g,  \label{31}
\end{equation}%
The fields $A,B_{mn},C_{n},\gamma _{mn}$ are neutral under this symmetry.
Then the current is invariant $\delta _{\varepsilon }J^{-m}=0.$ Furthermore,
in the action, the derivatives $\partial _{m}\varepsilon ^{MN}$ produced by
the two kinetic terms in (\ref{Ls}) cancel each other, while all other terms
involving $\varepsilon ^{MN}$ also cancel, so that $\delta _{\varepsilon
}\left( \sqrt{-\gamma }\mathcal{L}^{-}\right) =0$. Similarly, there is an SU$%
\left( 4\right) =$SO$\left( 6\right) $ local symmetry with parameters $%
\varepsilon ^{IJ}\left( \left( \tau ,\sigma \right) \right) $ under \textit{%
left multiplication} of $g$
\begin{equation}
\delta _{\omega }g=\frac{1}{4}\omega ^{IJ}\left(
\begin{array}{cc}
0 & 0 \\
0 & \Gamma _{IJ}%
\end{array}%
\right) \,g,  \label{32}
\end{equation}%
The fields $X,P^{m},A,B_{mn},C_{n},\gamma _{mn}$ are neutral under this
symmetry. The derivative $\partial _{m}\varepsilon ^{IJ}$ as well as other
dependence on $\varepsilon ^{IJ}$ drops both in the current and in the
action because $\Gamma _{IJ}$ and $\Gamma _{MN}$ appear in different blocks$%
. $

\item[3b-] Finally there is a local fermionic extended kappa (super)symmetry
under \textit{left multiplication} of $g$ with infinitesimal fermionic coset
elements $K\in $SU$\left( 2,2|4\right) _{L}$ of the form
\begin{equation}
\delta _{\kappa }g=Kg,\quad K=\left(
\begin{array}{cc}
0 & \xi \\
\tilde{\xi} & 0%
\end{array}%
\right) ,  \label{33}
\end{equation}%
provided $\delta _{\kappa }A,\delta _{\kappa }B_{mn},\delta _{\kappa }C_{n},$
are non-zero as specified below, and the local $\xi _{s}^{a}\left( \tau
,\sigma \right) $ has the form
\begin{equation}
\xi _{s}^{a}=X^{M}\left( \Gamma _{M}\kappa _{0}^{a}\right) _{s}+\left(
P^{-m}\right) ^{M}\left( \Gamma _{M}\kappa _{m}^{a}\right) _{s}\,\,,
\label{xi}
\end{equation}%
with the local fermionic parameters $\left( \kappa _{0}\right) _{s}^{a}$ and
$\left( \kappa _{-m}\right) _{s}^{a}$ unrestricted\footnote{%
However, since $X^{2}=P^{+2}=X\cdot P^{+}=0,$ the prefactors $X^{M}\Gamma
_{M}$ and $P^{+mM}\Gamma _{M}$ have zero eigenvalues. Therefore only part of
the kappa parameters can remove degrees of freedom from $g\left( \tau
,\sigma \right) $ by gauge fixing (e.g. as in Eq.(\ref{gfixed})). So there
remains physical fermionic degrees of freedom in $g.$}. The fields $%
X,P^{m},\gamma _{mn}$ are neutral under this symmetry. Then such a
transformation gives for the current $\delta _{\kappa }J^{-m}=ig^{-1}\left[
L^{-m},K\right] g,$ and for the action
\begin{equation}
\delta _{\kappa }\left( \sqrt{-\gamma }\mathcal{L}^{-}\right) =-\frac{1}{2}%
\delta _{\kappa }AX\cdot X-\frac{1}{2}\delta _{\kappa }B_{mn}P^{-m}\cdot
P^{-n}-\delta _{\kappa }C_{m}P^{-m}\cdot X+\frac{1}{2}Str\left( i\partial
_{m}gg^{-1}\left[ L^{-m},K\right] \right) ,  \label{dkappa}
\end{equation}%
with%
\begin{equation}
\left[ L^{-m},K\right] =X_{[M}P_{N]}^{-m}\left(
\begin{array}{cc}
0 & \Gamma ^{MN}\xi \\
-\tilde{\xi}\Gamma ^{MN} & 0%
\end{array}%
\right) .  \label{lkappa}
\end{equation}%
Inserting the general form in (\ref{xi}) we examine the product
\begin{equation}
X_{[M}P_{N]}^{-m}\left( \Gamma _{MN}\xi \right) =X_{[M}P_{N]}^{-m}\Gamma
_{MN}\left( X^{R}\left( \Gamma _{R}\kappa _{0}^{a}\right) _{A}+\left(
P^{-m}\right) ^{R}\left( \Gamma _{R}\kappa _{m}^{a}\right) _{A}\right) .
\label{34}
\end{equation}%
The three gamma term $\Gamma _{MNR}$ in the gamma matrix algebra $\Gamma
_{MN}\Gamma _{R}=\Gamma _{MNR}+\Gamma _{M}\eta _{NR}-\Gamma _{N}\eta _{MR}$
forces antisymmetry and drops out for any $\kappa _{0},\kappa _{-m}$. The
remaining one gamma terms give dot products $X\cdot X$, $P^{-m}\cdot P^{-n}$%
, $P^{-m}\cdot X$ , and those terms can be cancelled in the action by the
appropriate choice of $\delta _{\kappa }A$, $\delta _{\kappa }B_{mn}$, $%
\delta _{\kappa }C_{n}.$ In the case of the current we obtain $\delta
_{\kappa }J^{-m}=0$ on physical states when the vanishing of the quantities $%
X\cdot X$, $P^{-m}\cdot P^{-n}$, $P^{-m}\cdot X$ are applied as constraints
on physical states. Hence the action and the current are invariant under the
local kappa supersymmetry embedded in SU$\left( 2,2|4\right) _{L}$.
\end{itemize}

We can now specialize to some 1T cases of interest in $3+1$ dimensions by
using the gauge symmetries to thin out the degrees of freedom from $4+2$
dimensions to those in $3+1$ dimensions. This gives various holographic
pictures parallel to those discussed in the case of the particle in the
previous section. One gauge produces a superstring in 3+1 dimensions that
parallels the superparticle case of Eq.(\ref{L1}). Another one is the
twistor gauge that parallels Eq.(\ref{Ltwistor}), which we will discuss in
more detail in order to establish the twistor superstring as a gauge fixed
version of the 2T superstring.

We work in the conformal gauge $\sqrt{-\gamma }\gamma ^{mn}=\eta ^{mn}$
which reduces the system to the 2T string action in Eq.(\ref{Lss}).
Following the same arguments for the gauge choices that led to Eqs.(\ref%
{Ltwistor},\ref{JZZ}) for the particle, we fix the gauge for the string and
solve the constraints. This gives $X^{M}\left( \tau ,\sigma \right) =\left(
P^{-}\right) ^{M}\left( \tau ,\sigma \right) =0$ for all $\tau ,\sigma $ and
all $M$ except the following nonzero constant components for $M=+^{\prime
},- $
\begin{equation}
X^{+^{\prime }}\left( \tau ,\sigma \right) =1,\;\left( P^{-}\right)
^{-}\left( \tau ,\sigma \right) =1.  \label{XPfixed}
\end{equation}%
In this gauge the only nonzero component of $\left( L^{-}\right) ^{MN}$ is $%
\left( L^{-}\right) ^{+^{\prime }-}=1,$ or $L^{-}=\Gamma $ as in Eq.(\ref%
{JZZ}), and therefore the 2T string action and its conserved SU$\left(
2,2|4\right) _{R}$ current take the following gauge fixed forms%
\begin{eqnarray}
\mathcal{L}^{-} &=&-\frac{1}{4}Str\left( \partial _{-}gg^{-1}\Gamma \right) +%
\mathcal{L}_{1}^{-}=\bar{Z}^{A}\left( \partial _{-}Z_{A}\right) +\mathcal{L}%
_{1}^{-},  \label{Ltwstring} \\
J_{A}^{~B} &=&\left( \frac{1}{4}g^{-1}\Gamma g\right) _{A}^{B}=Z_{A}\bar{Z}%
^{B},\;  \label{Jstring}
\end{eqnarray}%
where $Z_{A}\left( \tau ,\sigma \right) $ are string supertwistors that
satisfy the constraint

\begin{equation}
J_{0}\equiv \bar{Z}^{A}Z_{A}=0,  \label{J0string}
\end{equation}%
which should be applied on physical states. The constraint arises from the
gauge symmetries of the 2T superstring as explained in the case of the
superparticle in the previous section. The twistor system that has emerged
in Eqs.(\ref{Ltwstring}-\ref{J0string}) is the same as the twistor
superstring version suggested by Berkovits \cite{berko,berko2}.

A new geometric description can also be given. As explained in \cite{witten}
the geometric space described by the constrained twistors is CP$^{3|4}.$ In
the 2T-physics approach we find that CP$^{3|4}$ is equivalent to the coset
space%
\begin{equation}
\text{CP}^{3|4}\leftrightarrow \text{PSU}\left( 2,2|4\right) /\text{H}%
_{\Gamma }
\end{equation}%
where H$_{\Gamma }$ is the subgroup of PSU$\left( 2,2|4\right) $ that
commutes with the constant matrix $\Gamma .$ This is the leftover gauge
symmetry after fixing $\left( X^{M},P^{M}\right) $ as in Eq.(\ref{XPfixed}).
The H$_{\Gamma }$ symmetry can remove further gauge degrees of freedom from $%
g$ and reduce it to the constrained supertwistors, i.e. CP$^{3|4}$. The Lie
superalgebra of H$_{\Gamma }$ is embedded in a ``triangular" configuration
in the $8\times 8$ supermatrix, and is given by
\begin{equation}
h_{\Gamma }=su\left( 1,1|4\right) +\text{V}_{\left( 1,1|4\right) }+\text{\={V%
}}_{\left( 1,1|4\right) }+R
\end{equation}%
V$_{\left( 1,1|4\right) }$ is in the fundamental representation of $su\left(
1,1|4\right) ,$ and \={V}$_{\left( 1,1|4\right) }$ is its complex conjugate.
Some of the supercommutators are $[su\left( 1,1|4\right) ,$V$_{\left(
1,1|4\right) }\}\sim $V$_{\left( 1,1|4\right) },$ and similarly for \={V}$%
_{\left( 1,1|4\right) }$. Another of the nontrivial supercommutators is $[$V$%
_{\left( 1,1|4\right) },$\={V}$_{\left( 1,1|4\right) }\}\sim R,$ while $R$
is an Abelian factor that commutes with all the other generators in $%
h_{\Gamma }.$ The counting of \textit{real} bosonic and fermionic parameters
is%
\begin{eqnarray}
\text{PSU}\left( 2,2|4\right) &:&\;bosons=~15+15,\;\;fermions=32 \\
su\left( 1,1|4\right) &:&\;bosons=~3+1+15,\;fermions=16 \\
\text{V}_{\left( 1,1|4\right) }+\text{\={V}}_{\left( 1,1|4\right) }
&:&\;bosons=~2+2,\;\;fermions=4+4 \\
R &:&\;bosons=1
\end{eqnarray}%
From this we see that the coset PSU$\left( 2,2|4\right) /$H$_{\Gamma }$ has $%
6$ real bosons and $8$ real fermions, which is the correct number of real
parameters in CP$^{3|4}.$ This is also the correct number\footnote{%
The physical phase space $\left( x,p,\theta \right) $ of the standard
one-time superparticle, in $d$ dimensions with $N$ supersymmetries, is $%
2(d-1)$ bosons and $N$ times half the dimension of the spinor representation
(if the spinor is complex multiply with another factor of 2).\label{superdim}%
} of physical phase space degrees of freedom $\left( x,p,\theta \right) $
for the superparticle given in Eq.(\ref{L1}), as it should be. From this
description of the geometry we see that we can present the supertwistor
string as a gauged sigma model with the global group SU$\left( 2,2|4\right) $
and the gauged subgroup H$_{\Gamma }.$

The arguments above show that the twistor superstring is a gauge fixed
version of the 2T superstring. Hence the quantization of the 2T superstring
in this gauge\footnote{%
A more complete BRST quantization would be to introduce the ghosts for all
the gauge symmetries SL$\left( 2,R\right) \times $SU$\left( 2,2\right)
\times $SU$\left( 4\right) \times Kappa$ where SU$\left( 2,2\right) \times $%
SU$\left( 4\right) \times Kappa$ $\subset $SU$\left( 2,2|4\right) .$ We
postpone this to a later investigation.} can be performed by following the
BRST quantization of the twistor system, including the appropriate ghosts,
as in \cite{berko,berko2}\cite{berkWit}.

Here we make some additional remarks regarding the SU$\left( 2,2|4\right)
_{R}$ symmetry that classifies the physical states of the system. From the
equations of motion it is evident that $\partial _{-}Z_{A}=0,$ so that $%
Z_{A} $ and $J_{A}^{B}$ are holomorphic as functions of $z=e^{i\left( \tau
+\sigma \right) }$.

The quantization rules for the twistors may be summarized by the operator
products%
\begin{equation}
Z_{A}\left( z\right) ~\bar{Z}^{B}\left( w\right) \sim \frac{\delta _{A}^{~B}%
}{z-w}  \label{wick}
\end{equation}%
Note that there is no pole in the sum $Z_{A}\left( z\right) ~\bar{Z}%
^{A}\left( w\right) \left( -1\right) ^{A}=\bar{Z}^{A}\left( w\right)
Z_{A}\left( z\right) $ due to the cancellation between an equal number of
bosons and fermions. Hence there is no problem of singularities as $%
z\rightarrow w$ in imposing the constraint $J_{0}\left( z\right) \equiv \bar{%
Z}^{A}\left( z\right) Z_{A}\left( z\right) \sim 0$ on physical states at the
quantum level.

The stress tensor is
\begin{equation}
T\left( z\right) =:\frac{1}{2}\partial _{z}\bar{Z}^{A}\left( z\right)
Z_{A}\left( z\right) -\frac{1}{2}\bar{Z}^{A}\left( z\right) \partial
_{z}Z_{A}\left( z\right) :+t\left( z\right)  \label{T}
\end{equation}%
where $t\left( z\right) $ comes from the $\mathcal{L}_{1}^{-}$ part of the
action. The dimensions of both $Z_{A}$ and $\bar{Z}^{A}$ is $1/2$ since they
are essentially hermitian conjugates of each other except for the SU$\left(
2,2|4\right) $ metric. Thus, as computed by using Eq.(\ref{wick}) we have
\begin{equation}
T\left( z\right) ~Z_{A}\left( w\right) \sim \frac{\frac{1}{2}Z_{A}\left(
w\right) }{\left( z-w\right) ^{2}}+\frac{\partial _{w}Z_{A}\left( w\right) }{%
\left( z-w\right) }
\end{equation}%
and similarly for $\bar{Z}^{A}.$ These dimensions are shifted from the
dimensions given in \cite{berko,berko2}\cite{berkWit} if we insist on
hermiticity with the spacetime signature for SO$(4,2)=$SU$\left( 2,2\right)
. $ However, for the analytic continuation of SO$\left( 4,2\right) $ to the
signature of SO$\left( 3,3\right) =$SL$\left( 4,R\right) ,$ and of SU$\left(
2,2|4\right) $ to SL$\left( 4|4;R\right) ,$ as used in \cite{berko,berko2}%
\cite{berkWit}, one may assign the dimensions $\dim \left( Z\right) =0$ and $%
\dim \left( \bar{Z}\right) =1.$ This amounts to a shift in the stress tensor
by a twist $T\rightarrow T_{0}\left( z\right) =T\left( z\right) -\frac{1}{2}%
\partial _{z}J_{0}=-\bar{Z}^{A}\left( z\right) \partial _{z}Z_{A}\left(
z\right) :+t\left( z\right) .$

For the computation of various SYM helicity amplitudes in nontrivial
instanton sectors one introduces further twisting to a stress tensor of
degree $n$ \cite{berko,berko2}\cite{berkWit}%
\begin{equation}
T_{n}\left( z\right) =T\left( z\right) -\frac{1}{2}\left( n+1\right)
\partial _{z}J_{0}=-\bar{Z}^{A}\partial _{z}Z_{A}-\frac{n}{2}\partial
_{z}\left( \bar{Z}^{A}Z_{A}\right) .  \label{Tn}
\end{equation}%
Relative to the twisted stress tensor the dimensions of $Z_{A}$ and $\bar{Z}%
^{A}$ are $-n/2$ and $1+n/2$ respectively%
\begin{equation}
T_{n}\left( z\right) Z_{A}\left( w\right) \sim \frac{-\frac{n}{2}Z_{A}\left(
w\right) }{\left( z-w\right) ^{2}}+\frac{\partial _{w}Z_{A}\left( w\right) }{%
\left( z-w\right) },\;\;T_{n}\left( z\right) \bar{Z}^{A}\left( w\right) \sim
\frac{\left( 1+\frac{n}{2}\right) \bar{Z}^{A}\left( w\right) }{\left(
z-w\right) ^{2}}+\frac{\partial _{w}\bar{Z}^{A}\left( w\right) }{\left(
z-w\right) },
\end{equation}%
as required in the SYM amplitude computations performed in \cite%
{berko,berko2}\cite{berkWit}.

By imposing the Virasoro and $J_{0}$ constraints on the physical states,
i.e. requiring dimension one vertex operators for the degree zero stress
tensor $T_{0}$ in Eq.(\ref{Tn}) $T_{0}\left( z\right) V\left( w\right) \sim
\frac{V\left( w\right) }{\left( z-w\right) ^{2}}+\frac{\partial _{w}V\left(
w\right) }{\left( z-w\right) }$, and $J_{0}\left( z\right) V\left( w\right)
\sim 0,$ one finds that only the states constructed with the lowest modes of
$Z_{A}\left( z\right) ,\bar{Z}_{A}\left( z\right) ,j^{a}\left( z\right) $
satisfy the physical state conditions. Hence the twistor superstring is
equivalent to a superparticle in the zero instanton sector. The only
physical states are then the SYM supermultiplet (supersingleton, as in the
previous section) in the adjoint representation of the group $G$ associated
with the current $j^{r},$ plus the conformal supergravity multiplet (which
contributes to loops). These are given by the dimension one vertex operators
of the form $V_{SYM}\left( z\right) =\phi _{r}\left( Z\left( z\right)
\right) j^{r}\left( z\right) $ and $V_{SUGRA}\left( z\right) =\bar{Z}%
^{A}\left( z\right) f_{A}\left( Z\left( z\right) \right) \oplus \partial
Z_{A}\left( z\right) g^{A}\left( Z\left( z\right) \right) $ respectively, as
explained in \cite{berko,berko2}\cite{berkWit}. In computations of SYM
helicity amplitudes, higher instanton sectors associated with $T_{n}$ are
needed; then the higher modes of the twistors contribute to those
computations \cite{berko,berko2}\cite{berkWit}.

The SU$\left( 2,2|4\right) _{R}$ current $J_{A}^{B}\left( z\right)
=:Z_{A}\left( z\right) \bar{Z}^{B}\left( z\right) :$ must be understood as
being normal ordered at the quantum level. It follows from Eq.(\ref{wick})
that it satisfies the standard operator products between supercurrents%
\begin{equation}
J_{A}^{~~B}\left( z\right) ~J_{C}^{~~D}\left( w\right) \sim \frac{\delta
_{A}^{~D}\delta _{C}^{~B}\left( -1\right) ^{BC+1}}{\left( z-w\right) ^{2}}+%
\frac{\left( -1\right) ^{BC}}{\left( z-w\right) }\left[ -J_{A}^{~D}\left(
w\right) ~\delta _{C}^{~B}+\left( -1\right) ^{A\left( B+C\right)
}J_{C}^{~B}\left( w\right) ~\delta _{A}^{~D}\right]
\end{equation}%
By taking the supertrace and using $\left( -1\right) ^{A}J_{A}^{~~A}=\bar{Z}%
^{A}Z_{A}=J_{0},$ we derive from the above the operator products%
\begin{equation}
J_{0}\left( z\right) J_{C}^{~~D}\left( w\right) \sim -\frac{\delta _{C}^{~D}%
}{\left( z-w\right) ^{2}},\;\;J_{A}^{~~B}\left( z\right) J_{0}\left(
w\right) \sim -\frac{\delta _{A}^{~B}}{\left( z-w\right) ^{2}}%
~,\;\;J_{0}\left( z\right) J_{0}\left( w\right) \sim 0  \label{JABJ0}
\end{equation}%
without any single poles. After subtracting the part of $J_{A}^{~~B}\left(
z\right) $ proportional to $\delta _{A}^{B},$ the remaining PSU$\left(
2,2|4\right) $ current has vanishing operator product with $J_{0}.$ We also
derive the matrix product of two SU$\left( 2,2|4\right) $ currents, obtained
by setting $B=C$ and summing. The coefficient of the double pole vanishes,
and we remain with
\begin{equation}
J_{A}^{~~B}\left( z\right) ~J_{B}^{~~D}\left( w\right) \sim \frac{%
J_{0}\left( w\right) ~\delta _{A}^{~D}}{\left( z-w\right) }.
\end{equation}%
Since $J_{0}\left( w\right) \sim 0$ on physical states, we see that the
matrix product of operators $\left( J\left( z\right) J\left( w\right)
\right) _{A}^{~~D}$ applied on physical states is finite as $z\rightarrow w.$

\section{2T superstring in 10+2\label{10+2}}

The 2T superparticle in $4+2$ dimensions discussed in a previous section was
generalized to the 2T superparticle in $10+2$ dimensions \cite{2tAdSs} by
adding six more bosons $\left( X^{I},P^{I}\right) ,$ $I=1,\cdots ,6,$ that
couple into the SO$\left( 6\right) =$SU$\left( 4\right) $ sector of SU$%
\left( 2,2|4\right) $. The global supersymmetry does not change, it is still
SU$\left( 2,2|4\right) _{R},$ and it classifies the physical states. One of
the gauge fixed versions of this 2T particle model gives the AdS$_{5}\times $%
S$^{5}$ superparticle. Its quantum states were computed and summarized in
Eq.(4.29) and footnote (4) of ref.\cite{2tAdSs}. The spectrum matches to the
well known Kaluza-Klein towers of type-IIB supergravity compactified on AdS$%
_{5}\times $S$^{5}.$

In this section we develop the parallel formalism for the 2T superstring in $%
10+2$ dimensions. We will then choose a 1T gauge (equivalent to a left
moving AdS$_{5}\times $S$^{5}$ string) that reduces the theory to a
collection of 8 supertwistors and their conjugates$,$ subject to a set of
constraints that satisfy the S(U$\left( 2|2\right) \times $U$\left(
2|2\right) $) algebra. The constrained twistors describe the degrees of
freedom in the coset space SU$\left( 2,2|4\right) /$S(U$\left( 2|2\right)
\times $U$\left( 2|2\right) $). This space contains 8 complex bosons and 8
complex fermions which are equivalent to the physical phase space of AdS$%
_{5}\times $S$^{5}$ string in a lightcone gauge. The well understood
particle limit spectrum suggests an SU$\left( 2,2|4\right) _{R}$
classification of string states in compactified 10+2 dimensions (F-theory or
S-theory). It is found that this spectrum corresponds to the conserved high
spin currents expected in the weak coupling limit of $N=4,$ $d=4$ SYM
theory, in agreement with AdS-CFT correspondence.

The action of the 2T superstring in 10+2 dimensions has the same form as $%
\mathcal{L}^{-}$ in Eq.(\ref{Ls}) but instead of six dimensions there are
twelve dimensions labelled as $\hat{X}^{\hat{M}}=\left( X^{M},X^{I}\right) $
and $\hat{P}^{-\hat{M}}=\left( P^{-M},P^{-I}\right) ,$ with $I=1,\cdots ,6,$
$M=+^{\prime },-^{\prime },\mu ,$ and $\mu =0,1,2,3.$ The six dimensions
labelled with $M$ are the same as those of the $4+2$ string, while the extra
six dimensions labelled by $I$ appear as their extension into 12 dimensions.
Written in the conformal gauge as in Eq.(\ref{Lss}) the Lagrangian takes the
form
\begin{equation}
\sqrt{-\gamma }\mathcal{\hat{L}}^{-}=\partial _{-}\hat{X}\cdot \hat{P}^{-}-%
\frac{1}{2}A\hat{X}\cdot \hat{X}-\frac{1}{2}B_{--}\hat{P}^{-}\cdot \hat{P}%
^{-}-C_{-}\hat{P}^{-}\cdot \hat{X}+\frac{1}{2}Str\left( i\partial _{-}gg^{-1}%
\hat{L}^{-}\right) +\mathcal{L}_{1}^{-}.  \label{L12}
\end{equation}%
with
\begin{equation}
\hat{L}^{-}\equiv \left(
\begin{array}{cc}
\frac{i}{2}\Gamma ^{MN}X_{[M}P_{N]}^{-} & 0 \\
0 & -\frac{i}{2}\Gamma ^{IJ}X_{[I}P_{J]}^{-}%
\end{array}%
\right) .
\end{equation}%
The extra dimensions appear in the 12-dimensional dot products $\hat{X}\cdot
\hat{X},\hat{P}^{-}\cdot \hat{P}^{-},\hat{P}^{-}\cdot \hat{X}$ and in the
second block of $\hat{L}^{-}.$ The dot products are invariant under SO$%
\left( 10,2\right) $ but $Str\left( i\partial _{-}gg^{-1}\hat{L}^{-}\right) $
reduces the symmetry to the subgroup SO$\left( 4,2\right) \times $SO$\left(
6\right) .$ The extra minus sign in the lower block of $\hat{L}^{-}$ is
cancelled by the extra minus sign in the supertrace.

The global and local symmetries are similar to those listed in Eqs.(\ref{gR}-%
\ref{34}) with slight modifications that are outlined below. The most
important modification is the kappa supersymmetry as discussed in item 3
below.

\begin{enumerate}
\item[1-] The right side symmetry SU$\left( 2,2|4\right) _{R}$ acts as in
Eq.(\ref{gR}), and has a conserved holomorphic current $\hat{J}^{-}=\frac{1}{%
4}g^{-1}\hat{L}^{-}g,\;$just as before, but with $\hat{L}^{-}$ replacing $%
L^{-}.$

\item[2-] The local Sp$\left( 2,R\right) $ symmetry parameterized by $\alpha
,\beta ,\rho $ act on all 12 doublets $\left( \hat{X}^{\hat{M}},\hat{P}^{-%
\hat{M}}\right) $ and the gauge fields $\left( A,B_{--},C_{-}\right) $ as in
Eqs.(\ref{21}-\ref{23}). Both blocks of $\hat{L}^{-}$ are separately
invariant under Sp$\left( 2,R\right) .$ The SO$\left( 10,2\right) $
invariant Sp$\left( 2,R\right) $ constraints $\hat{X}\cdot \hat{X}=\hat{P}%
^{-}\cdot \hat{P}^{-}=\hat{P}^{-}\cdot \hat{X}$ $=0$ include all 12
dimensions, and not each six dimensional subset separately. Note that the
solution space of these constraints include the case of vanishing extra
dimensions $X_{I}=P_{I}^{-}=0$ (i.e. $4+2$ theory recovered as a special
solution).

\item[3-] The local SU$\left( 2,2|4\right) _{L}$ symmetry acts on $g$ as in
Eqs.(\ref{31}-\ref{33}), on $\left( X^{M},P^{-M}\right) $ as in Eq.(\ref{31}%
), and on $\left( X^{I},P^{-I}\right) $ as $\delta _{\omega }X^{I}=\omega
^{IJ}X_{J},~\delta _{\omega }\left( P^{-}\right) ^{J}=\omega ^{IJ}\left(
P_{J}^{-}\right) .$ Hence the SO$\left( 4,2\right) =$SU$\left( 2,2\right) $
and SO$\left( 6\right) =$SU$\left( 4\right) $ are local symmetries. The
kappa supersymmetry in Eqs.(\ref{xi}-\ref{34}) is modified as follows. The
form of $\xi _{s}^{a}$ in $K$ remains the same for the special solution
subspace when the Sp$\left( 2,R\right) $ constraints are satisfied with
vanishing $\left( X^{I},P^{-I}\right) $, but otherwise is modified such
that, instead of Eq.(\ref{xi}), we now have (still Sp$\left( 2,R\right) $
invariant $L^{-MI}$ or $\xi _{s}^{a}$)
\begin{equation}
\xi _{s}^{a}=L^{-MI}\,\left( \Gamma _{M}\kappa \Gamma _{I}\right)
_{s}^{a}=X^{M}\left( \Gamma _{M}\kappa \Gamma _{I}\right)
_{s}^{a}P^{-I}-P^{-M}\left( \Gamma _{M}\kappa \Gamma _{I}\right)
_{s}^{a}X^{I}.  \label{xi2}
\end{equation}%
There is only one free parameter $\kappa _{s}^{a}$ instead of the two in Eq.(%
\ref{xi}). Then the $\delta _{\kappa }$ transformation of the current $%
\delta _{\kappa }J$ and the action $\delta _{\kappa }\mathcal{\hat{L}}^{-}$
produce terms involving $\left[ \hat{L}^{-},K\right] $ as in Eqs.(\ref%
{dkappa},\ref{lkappa}) but $\left[ \hat{L}^{-},K\right] $ now has the form%
\begin{equation}
\left[ \hat{L}^{-},K\right] \sim \left(
\begin{array}{cc}
0 & L^{MN}\left( \Gamma _{MN}\xi \right) +\left( \xi \Gamma _{IJ}\right)
L^{IJ} \\
-L^{IJ}\left( \Gamma _{IJ}\bar{\xi}\right) -\left( \bar{\xi}\Gamma
_{MN}\right) L^{MN} & 0%
\end{array}%
\right) .  \label{LK}
\end{equation}%
When the $\xi $ in Eq.(\ref{xi2}) is inserted in the structure $L^{MN}\left(
\Gamma _{MN}\xi \right) +L^{IJ}\left( \xi \Gamma _{IJ}\right) ,$ the three
gamma terms $\Gamma ^{MNR}$ or $\Gamma ^{IJK}$ force antisymmetry, and drop
out, while the remainder is seen to reduce to a linear combination of the SO$%
\left( 10,2\right) $ covariant dot products $\hat{X}\cdot \hat{X},~\hat{P}%
^{-}\cdot \hat{P}^{-},~\hat{P}^{-}\cdot \hat{X}$. Therefore, these can be
cancelled in $\delta _{\kappa }\mathcal{\hat{L}}^{-}$ (see the form in Eq.%
\ref{dkappa}) by choosing $\delta _{\kappa }A,~\delta _{\kappa
}B_{--},~\delta _{\kappa }C_{-}.$ Similarly, in the kappa variation of the
current $\delta _{\kappa }J$ the Sp$\left( 2,R\right) $ constraints vanish
on physical states, so that $\delta _{\kappa }J\sim 0$ on Sp$\left(
2,R\right) $ gauge invariant physical states. Hence there is a kappa
supersymmetry in the 10+2 theory embedded in SU$\left( 2,2|4\right) _{L}.$
\end{enumerate}

It should be noted that the kappa supersymmetry is smaller when the Sp$%
\left( 2,R\right) $ constraints are satisfied with nonvanishing $L^{IJ}.$
Then there is a single kappa parameter instead of two as noted above. This
amount of kappa supersymmetry can remove only half of the fermions in $%
g\left( \tau ,\sigma \right) $. By contrast, for the special solution for
which the lower block of $\hat{L}^{-}$ vanishes, the larger kappa
supersymmetry can remove 3/4 of the fermions in $g$ as in the $4+2$ string
of the previous section. The amount of kappa supersymmetry has a profound
effect on the physical spectrum. With 1/2 kappa supersymmetry, as in the
generic solutions of the 10+2 theory, the physical spectrum (at the particle
limit) is supergravity, while with 3/4 kappa supersymmetry, as in the 4+2
theory or the equivalent special solution of the 10+2 theory, the physical
spectrum (at the particle limit) is SYM theory.

We now examine the physical content of this theory by choosing some 1T
gauges. The AdS$_{5}\times $S$^{5}$ gauge is obtained by fixing two Sp$%
\left( 2,R\right) $ gauges $\left( P^{-}\right) ^{+^{\prime }}\left( \tau
,\sigma \right) =0$, $\left\vert X^{I}\right\vert \left( \tau ,\sigma
\right) =R=$constant, and solving two of the Sp$\left( 2,R\right) $
constraints $\hat{X}\cdot \hat{X}=\hat{X}\cdot \hat{P}^{-}=0.$ The resulting
phase space takes the form
\begin{align}
\hat{M}& =\left( \,\,+^{\prime }\quad \quad \,\,\,-^{\prime }\quad \quad
\,\quad \,\mu \quad \quad \,\quad I\right)  \notag \\
\hat{X}^{\hat{M}}\left( \tau ,\sigma \right) & =\frac{R}{\left\vert \mathbf{y%
}\right\vert }\left( R,\,\,\,\,\,\,\,\frac{x^{2}+\mathbf{y}^{2}}{2R},\quad
x^{\mu }\,,\quad \mathbf{y}^{I}\right) \left( \tau ,\sigma \right)
\label{adsX} \\
\hat{P}^{\hat{M}}\left( \tau ,\sigma \right) & =\frac{\left\vert \mathbf{y}%
\right\vert }{R}\left( 0,\,\,\,\,\frac{1}{R}\left( x\cdot p+\mathbf{y\cdot k}%
\right) \,,\,\,\,\,\,\,\,\,\,p^{\mu }\,\,\,,\,\,\mathbf{\,\,\,k}^{I}\right)
\left( \tau ,\sigma \right) .  \label{adsP}
\end{align}%
Evidently, we obtain $X^{M}X_{M}=-R^{2}$ and $X^{I}X_{I}=R^{2}$ which is the
AdS$_{5}\times $S$_{5}$ space given by the metric%
\begin{equation}
ds^{2}=d\hat{X}^{\hat{M}}d\hat{X}_{\hat{M}}=\frac{R^{2}}{y^{2}}\left[ \left(
dx^{\mu }\right) ^{2}+\left( dy\right) ^{2}\right] +\left( d\mathbf{\Omega }%
\right) ^{2}.  \label{metric}
\end{equation}%
where $\mathbf{\Omega }^{I}=\mathbf{y}^{I}/\left\vert \mathbf{y}\right\vert $
and $y=\left\vert \mathbf{y}\right\vert .$ The boundary of the AdS$_{5}$
space at $y\rightarrow 0$ is Minkowski space $x^{\mu }$ in 4-dimensions. The
SU$\left( 2,2|4\right) _{L}$ gauge symmetry can be used to gauge fix $g$ to
the form Eq.(\ref{gfixed}). In this gauge the $10+2$ superstring reduces to
a left moving AdS$_{5}\times $S$^{5}$ superstring.

The particle limit of this theory was analyzed in this gauge, and its
quantum spectrum was summarized in Eq.(4.29) and footnote (4) of ref.\cite%
{2tAdSs}. The particle limit spectrum is $2^{7}$ bosons and $2^{7}$ fermions
with AdS$_{5}\times $S$^{5}$ quantum numbers
\begin{equation}
\Phi _{2_{B}^{7}+2_{F}^{7}}\left( x^{\mu },y,l\right) Y_{l}\left( \Omega
\right)  \label{KK}
\end{equation}%
where $Y_{l}\left( \Omega \right) ,~l=0,1,2,\cdots $ is a symbol for
harmonics on S$^{5}$ (one row symmetric rank $l$ traceless tensors of SO$%
\left( 6\right) $ constructed from the vector $\Omega ^{I}$). These states
satisfy the 12-dimensional mass shell condition $\hat{P}^{2}=0,$ which in
this gauge takes the form $\Delta _{AdS_{5}}\Phi
_{2_{B}^{7}+2_{F}^{7}}\left( x^{\mu },y,l\right) =l\left( l+4\right) \Phi
_{2_{B}^{7}+2_{F}^{7}}\left( x^{\mu },y,l\right) .$ The $l=0$ case is the
special solution that reduces to the $4+2$ superparticle, which has a larger
kappa symmetry. Therefore, for $l=0$ the spectrum reduces to the short
supermultiplet with $2^{3}$ bosons plus $2^{3}$ fermions, which gives the
SYM supermultiplet, as already discussed earlier in this paper. For general $%
l$, since the model has a global symmetry SU$\left( 2,2|4\right) _{R}$ the
states are classified as towers of SU$\left( 2,2\right) $ distinguished by
the SU$\left( 4\right) = $SO$\left( 6\right) $ quantum number $l.$ It was
shown that for $l\geq 1$ this is the same as the Kaluza-Klein spectrum of
linearized type-IIB supergravity compactified on AdS$_{5}\times $S$^{5},$
while for $l=0$ it is the singleton equivalent to the $N=4,$ $d=4$ SYM
spectrum, as discussed earlier in this paper.

The results of the particle case described in the previous paragraph suggest
that the string case of the present paper generalizes the compactified
type-IIB supergravity spectrum to a compactified string spectrum on AdS$%
_{5}\times $S$^{5},$ and furthermore that this spectrum should be organized
as representations of the current algebra SU$\left( 2,2|4\right) _{R}$ since
this is the global symmetry of the theory. There remains to study the
representations of this non-compact super current algebra or use related
methods (such as supertwistors as described below) to study the spectrum and
further properties of the theory.

Before plunging into detailed computation it is interesting to note that
there is a candidate spectrum that was suggested in 1995 on the basis of
symmetries in M-theory \cite{11Dstring} and was recently revived in the
context of $N=4,$ $d=4$ SYM theory and the AdS-CFT correspondence \cite%
{bianchi}. This provides a useful guide to organize the spectrum we are
seeking, to relate it to other interesting concepts, and to simultaneously
use the newly emerging framework as a basis for the group theoretical
classification found in \cite{11Dstring}\cite{bianchi}.

In 1995 it was suggested that the massive 10D type-IIA string spectrum could
be extended to compactified 11D M-theory massive spectrum, including
Kaluza-Klein (KK) states, just like the massless 10D type-IIA spectrum is
extended to compactified 11D supergravity spectrum. The guiding tool was the
little group SO$\left( 10\right) $ for massive states in 11D, and one needed
to find the completion of the SO$\left( 9\right) $ massive string spectrum
into SO$\left( 10\right) $ representations SO$\left( 9\right) \subset $SO$%
\left( 10\right) \subset $SO$\left( 10,1\right) $. In this way a systematic
formula for the spectrum including Kaluza-Klein states was discovered at all
string levels. The formula given in Eq.(3.8) in \cite{11Dstring} is very
simple. Define the total level $n$ as the string level $n-k$ plus the KK
level $k$. Start with the left-moving states at string mass level $n$ at KK
level $0$, then add the left-moving string states of level $n-1$ at KK level
$1$, plus those of string level $n-2$ at KK level $2$, and so on, up to the
left-moving string states of level $1$ at KK level $n-1.$ Repeat the same
procedure for the right moving sector at total level $n$, and then take the
product of left$\times $right movers each with total level $n$. The
collection of these states form SO$\left( 10\right) $ multiplets for every
total level $n\geq 1.${} The SO$\left( 10\right) $ representations obtained
in this way were given explicitly up to total level $n=5$ in \cite{11Dstring}%
.

To apply this formula to the present case, recall that the T-dual of
type-IIA is type-IIB. As long as one discusses the little group SO$\left(
9\right) \subset $SO$\left( 9,1\right) $ of the string, there is no
difference between starting with SO$\left( 9\right) $ representations of
type-IIA or type-IIB strings. The higher dimensional extension of type-IIB
is F-theory \cite{Ftheory} or S-theory \cite{Stheory}\cite{liftM} in 10+2
dimensions, with the compact subgroup SO$\left( 10\right) .$ The SO$\left(
10\right) $ classification of states described in the previous paragraph can
be interpreted (via T-duality) as those of a 10+2 dimensional theory. The
10+2 superstring suggested in this paper is expected to have a closely
related spectrum after compactification of $10+2$ to $\left( 4+2\right)
+\left( 6+0\right) $, with SO$\left( 10\right) \rightarrow $SO$\left(
4\right) \times $SO$\left( 6\right) .$ Indeed, we have already argued in
this section that the 10+2 string can be viewed as a 1T string on an AdS$%
_{5}\times $S$_{5}$ background, and that its particle limit produces the
compactified type-IIB supergravity spectrum. Therefore, to compare the
spectrum of \cite{11Dstring} to the present case, the SO$\left( 10\right) $
multiplets given in \cite{11Dstring} should be decomposed into SO$\left(
4\right) \times $SO$\left( 6\right) .$ Furthermore the SO$\left( 6\right) $
quantum numbers coming from the harmonic expansion of higher dimensional
fields into Kaluza-Klein towers should be included, as in Eq.(\ref{KK}).
These towers should then produce a series of SU$\left( 2,2|4\right) $
representations that can be compared to the SU$\left( 2,2|4\right) _{R}$
current algebra spectrum we are seeking.

Based on AdS-CFT correspondence we might expect that the infinite $N=4,$ $%
d=4 $ SYM theory has a close relationship with the spectrum produced by our
10+2 string taken in the AdS$_{5}\times $S$^{5}$ gauge described above. In
particular, we already know that the $10+2$ string has the special solution
sector of the $4+2$ string, which is indeed related to $N=4,$ $d=4$ SYM
theory through the twistor superstring, as shown earlier in this paper. As
further evidence we note that the SO$\left( 10\right) \rightarrow $SO$\left(
4\right) \times $SO$\left( 6\right) $ reduction process described in the
previous paragraph has precisely the same content of towers of SO$\left(
4\right) \times $SO$\left( 6\right) $ representations as the classification
of high spin currents \cite{bianchi} expected in the weak coupling limit
\cite{wittenW} of $N=4,$ $d=4$ SYM theory. This is encouraging for our
expected results on the spectrum of the $10+2$ string. The work in \cite%
{bianchi} parallels the group theoretical steps in \cite{11Dstring}, while
the current paper provides a dynamical string model with the same group
theoretical properties and with connections to SYM. The conclusive analysis
of the dynamics and of the group theory could be achieved through the
twistor framework emerging in the current paper, as described in the
remainder of this section.

To analyze the 10+2 superstring we now choose a twistor gauge instead of the
AdS$_{5}\times $S$^{5}$ gauge of Eqs.(\ref{adsX},\ref{adsP}). This is the
analog of the twistor gauge of Eq.(\ref{Ltwistor}) instead of the particle
gauge of Eq.(\ref{L1}) for the superparticle. Thus, we first use the local SO%
$\left( 4,2\right) \times $SO$\left( 6\right) \subset $SU$\left(
2,2|4\right) _{L}$ to rotate the 12 components of $\hat{X}^{\hat{M}}\left(
\tau ,\sigma \right) $ so that they point in the special directions $%
M=0^{\prime }$ and $I=1,$ and also impose the constraint $\hat{X}\cdot \hat{X%
}=0$. The result is the 12-dimensional lightlike vector $\hat{X}^{\hat{M}%
}\sim \left( 1,0,0,0,0,0;1,0,0,0,0,0\right) $, assuming that we are
analyzing the $L^{IJ}$ non-zero sector of the theory (i.e. not the $4+2$
special solution, which is already discussed in the previous section). There
still remains local symmetry SO$\left( 4,1\right) \times $SO$\left( 5\right)
\subset $SO$\left( 4,2\right) \times $SO$\left( 6\right) $ which does not
change the gauged fixed form of $\hat{X}$. Using this we can rotate $P^{-M}$
and $P^{-I}$ to special directions with at least four zero components each.
Then, using the Sp$\left( 2,R\right) $ local symmetry some of the remaining
non-zero components can be rotated to zero. Finally, applying the remaining
constraints $\hat{X}\cdot \hat{P}^{-}=\hat{P}^{-}\cdot \hat{P}^{-}=0$ we can
complete the gauge fixing of the 12 dimensional phase space $\hat{X}^{\hat{M}%
},\hat{P}^{-\hat{M}}$ to the form of two lightlike orthogonal vectors in 12
dimensions
\begin{align}
\hat{M}& =\left( \,\,0^{\prime }~~0~~~1~\cdots ~4~~,~I=1\,~2~~3~~\cdots
~6\right)  \notag \\
\hat{X}^{\hat{M}}\left( \tau ,\sigma \right) & \sim \left(
\,\,1~~~0~~~0~~\cdots ~0~~,~~~~1\,~~~0~~~~0~~\cdots ~0\right) \\
\hat{P}^{\hat{M}}\left( \tau ,\sigma \right) & \sim \left(
\,\,0~~~1~~~0~~\cdots ~0~~,~~~~0\,~~~1~~~~0~~\cdots ~0\right)
\end{align}%
In this gauge the $8\times 8$ matrix $\hat{L}^{-}$ simplifies to%
\begin{equation}
\hat{L}^{-}\sim \left(
\begin{array}{cc}
i\Gamma _{0^{\prime }0} & 0 \\
0 & i\Gamma _{12}%
\end{array}%
\right) \equiv \hat{\Gamma}=\left(
\begin{array}{cccc}
1_{2} & 0 & 0 & 0 \\
0 & -1_{2} & 0 & 0 \\
0 & 0 & -1_{2} & 0 \\
0 & 0 & 0 & 1_{2}%
\end{array}%
\right)
\end{equation}%
since the only nonzero components of $L^{MN},L^{IJ}$ are $L^{0^{\prime }0}$,
$L^{12}$ respectively, and furthermore $L^{0^{\prime }0}=L^{12}.$ We chose a
particular basis of gamma matrices so that $i\Gamma _{0^{\prime }0}$ and $%
i\Gamma _{12}$ are diagonal as written. Note that $\hat{\Gamma}$ is an
invariant under H$_{\hat{\Gamma}}=$S$\left( \text{U}\left( 2|2\right) \times
\text{U}\left( 2|2\right) \right) $ transformations embedded in SU$\left(
2,2|4\right) .$ Therefore, H$_{\hat{\Gamma}}\subset $SU$\left( 2,2|4\right)
_{L}$ is a remaining local symmetry that can remove further degrees of
freedom from the group element $g\left( \tau ,\sigma \right) .$ The first SU$%
\left( 2|2\right) $ acts on rows $R=1,2,7,8$ (labelled as $r$ below) and the
second SU$\left( 2|2\right) $ acts on rows $R=3,4,5,6$ (labelled as $%
r^{\prime }$ below) as seen from the form of $\hat{\Gamma}.$ The remaining U$%
\left( 1\right) $ has a generator proportional to $\hat{\Gamma}.$

In this gauge the action in Eq.(\ref{L12}) and the SU$\left( 2,2|4\right)
_{R}$ symmetry current reduce to%
\begin{eqnarray}
\sqrt{-\gamma }\mathcal{\hat{L}}^{-} &=&\frac{1}{4}Str\left( \partial
_{-}gg^{-1}\hat{\Gamma}\right) +\mathcal{L}_{1}^{-}=\bar{Z}_{r}^{A}\partial
_{-}Z_{A}^{r}\left( -1\right) ^{r}-\bar{Z}_{r^{\prime }}^{A}\partial
_{-}Z_{A}^{r^{\prime }}\left( -1\right) ^{r^{\prime }}+\mathcal{L}_{1}^{-}
\label{Lhattwist} \\
\left( \hat{J}^{-}\right) _{A}^{~B} &=&\left( \frac{1}{4}g^{-1}\hat{\Gamma}%
g\right) _{A}^{~B}=Z_{A}^{r}\bar{Z}_{r}^{B}-Z_{A}^{r^{\prime }}\bar{Z}%
_{r^{\prime }}^{B}  \label{Jhattwist}
\end{eqnarray}%
Thus, the theory is described by a collection of 8 supertwistors and their
conjugates, $Z_{A}^{~r}$,$\bar{Z}_{r}^{~A},$ $r=1,2,7,8,$ and $%
Z_{A}^{~r^{\prime }}$,$\bar{Z}_{r^{\prime }}^{~A},$ $r^{\prime }=3,4,5,6.$
The twistors labelled by $r=1,2$ and $r^{\prime }=3,4$ have bosons in their
first four components $A=1,2,3,4$ (basis for SU$\left( 2,2\right) \subset $SU%
$\left( 2,2|4\right) _{R}$) and fermions in their last four components $%
A=5,6,7,8$ (basis for SU$\left( 4\right) \subset $SU$\left( 2,2|4\right)
_{R} $). By contrast, the twistors labelled by $r=7,8$ and $r^{\prime }=5,6$
are unusual since they have fermions in the SU$\left( 2,2\right) $ basis and
bosons in the SU$\left( 4\right) $ basis. This structure is dictated by the
fact that, combined together, they make up the group element $g.$ The
raising or lowering of the indices on the twistors and their conjugates is
done in accordance with the fact that $g^{-1}$ is constructed by taking the
hermitian conjugate, and multiplying with the SU$\left( 2,2|4\right) $
metric, $g^{-1}=\eta g^{\dagger }\eta $. Therefore, the twistors are
constrained by the condition $g^{-1}g=1,$ which requires%
\begin{equation}
\left( j\right) _{r_{1}}^{~r_{2}}\equiv \bar{Z}_{r_{1}}^{~A}Z_{A}^{~r_{2}}=%
\delta _{r_{1}}^{r_{2}},\;\;\left( j^{\prime }\right) _{r_{1}^{\prime
}}^{~r_{2}^{\prime }}\equiv \bar{Z}_{r_{1}^{\prime
}}^{~A}Z_{A}^{~r_{2}^{\prime }}=\delta _{r_{1}^{\prime }}^{r_{2}^{\prime
}},\;\;\bar{Z}_{r}^{~A}Z_{A}^{~r^{\prime }}=\bar{Z}_{r^{\prime
}}^{~A}Z_{A}^{~r}=0.
\end{equation}%
These may be understood as arising from the remaining gauge symmetry
\begin{equation}
\text{H}_{\hat{\Gamma}}=\text{S}\left( \text{U}\left( 2|2\right) \times
\text{U}\left( 2|2\right) \right) _{L}\subset \text{SU}\left( 2,2|4\right)
_{L}
\end{equation}%
These constraints also guarantee that the SU$\left( 2,2|4\right) _{R}$
current $\left( \hat{J}^{-}\right) _{A}^{~B}$ has zero supertrace%
\begin{eqnarray}
Str\left( \hat{J}^{-}\right) &=&\left( Z_{A}^{r}\bar{Z}_{r}^{A}-Z_{A}^{r^{%
\prime }}\bar{Z}_{r^{\prime }}^{A}\right) \left( -1\right) ^{A}=\bar{Z}%
_{r}^{A}Z_{A}^{r}\left( -1\right) ^{r}-\bar{Z}_{r^{\prime
}}^{A}Z_{A}^{r^{\prime }}\left( -1\right) ^{r^{\prime }} \\
&=&Str\left( j\right) -Str\left( j^{\prime }\right) =Str\left( 1\right)
-Str\left( 1\right) =0
\end{eqnarray}%
where the patterns of signs $\left( -1\right) ^{r},\left( -1\right)
^{r^{\prime }},\left( -1\right) ^{A}$ take into account the interchange of
orders of bosons and fermions, and the definition of supertrace.

The quantization of this twistor system is given by the operator products%
\begin{equation}
Z_{A}^{r_{1}}\left( z\right) \bar{Z}_{r_{2}}^{B}\left( w\right) \sim \frac{%
\delta _{A}^{B}\delta _{r_{2}}^{r_{1}}}{z-w},\;\;Z_{A}^{r_{1}^{\prime
}}\left( z\right) \bar{Z}_{r_{2}^{\prime }}^{B}\left( w\right) \sim -\frac{%
\delta _{A}^{B}\delta _{r_{2}^{\prime }}^{r_{1}^{\prime }}}{z-w}%
,\;\;Z_{A}^{r}\left( z\right) \bar{Z}_{r^{\prime }}^{B}\left( w\right) \sim 0
\end{equation}%
Equivalently, we can write for the group element%
\begin{equation}
\left( g^{-1}\right) _{A}^{~R_{1}}\left( z\right) ~\left( g\right)
_{R_{2}}^{~B}\left( w\right) \sim \frac{\delta _{A}^{B}~\Gamma
_{R_{2}}^{R_{1}}}{z-w},~
\end{equation}%
with the current $\left( \hat{J}^{-}\right) _{A}^{~B}=\left( \frac{1}{4}%
g^{-1}\Gamma g\right) _{A}^{~B}$ and the constraints as given above.

These quantization rules are equivalent to a system of bosonic and fermionic
oscillators which are constrained as indicated. From the point of view of
the oscillator formalism for non-compact supergroups \cite{barsgunaydin}\cite%
{2tZero} the system can be interpreted as oscillators with a `` color"\
supergroup S$\left( \text{U}\left( 2|2\right) \times \text{U}\left(
2|2\right) \right) _{L}.$ The physical states are the `` color"\ singlets.
Thus, the representation space of the physical currents SU$\left(
2,2|4\right) _{R}$ can be analyzed with the kinds of algebraic oscillator
methods used in the past, after taking into account the fact that the ``
color"\ group is now a supergroup, and restricting the Fock space states to
the ``color" singlet sector as in \cite{2tZero}.

It should also be noted that geometric methods based on the coset
\begin{equation}
\text{SU}\left( 2,2|4\right) /\text{S}\left( \text{U}\left( 2|2\right)
\times \text{U}\left( 2|2\right) \right)
\end{equation}%
could provide another useful approach for analyzing the theory. One may
introduce the gauge fields of S$\left( \text{U}\left( 2|2\right) \times
\text{U}\left( 2|2\right) \right) $ explicitly to present the twistor model
above as a gauged sigma model$.$ This coset contains 2 real bosons plus 8
complex bosons, and 8 complex fermions (i.e. 18 real bosons and 16 real
fermions). This counting of independent degrees of freedom is the same as
the super phase space (positions, momenta and fermions) of the AdS$%
_{5}\times $S$^{5}$ string of Eqs.(\ref{adsX},\ref{adsP}) after fixing a
physical gauge. This makes it evident that the constrained twistor space
given above is equivalent to the conventional description in usual
spacetime, as expected from the fact that they are both obtained by gauge
fixing the same 10+2 superstring. In particular, the fermionic zero modes of
this coset create $2^{7}$ bosons and $2^{7}$ fermions. Given the SU$\left(
2,2|4\right) $ symmetry of the model, it is evident that these are the
correct states that describe the compactified supergravity multiplet, as
expected from Eq.(\ref{KK}).

The technical analysis of the twistor system above is incomplete at this
stage. We hope to discuss it in a future paper.

\section{Further remarks on other dimensions}

The 2T superstring in 4+2 dimensions is directly generalized to $d+2$
dimensions for the special dimensions $d=3,4,5,6$ by taking $\left(
X^{M},P^{mM}\right) \left( \tau ,\sigma \right) $ in the corresponding $d+2$
dimensions, and using the supergroup element $g\left( \tau ,\sigma \right)
\in G,$ with $G$ given by OSp$\left( 8|4\right) ,$ SU$\left( 2,2|4\right) ,$
F$\left( 4\right) ,$ OSp$\left( 8^{\ast }|4\right) $ respectively for $%
d=3,4,5,6$. The 2T Lagrangian has the same form as (\ref{Ls}), except for
modifying $L,L^{-m}$ in Eqs.(\ref{L},\ref{Lm}) by replacing $\frac{i}{2}%
\Gamma _{MN}L^{MN}\rightarrow i\frac{2}{s}\Gamma _{MN}L^{MN}$, where $s$ is
the dimension of the spinor in $d+2$ dimensions. The conserved current is as
before $J=\frac{1}{2}g^{-1}Lg$ for the group $G,$ and the local symmetries
are exact parallels as the ones discussed in items $1,2a,2b,2c,3a,3b$ in
section (\ref{4+2}). The normalization $i\frac{2}{s}\Gamma _{MN}L^{MN}$ is
needed to insure the SO$\left( d,2\right) $ local symmetry in item $3a.$

Just like the $d=4$ case, in the particle limit in each one of these string
models for $d=3,4,5,6,$ the physical states consist of 8 bosons and 8
fermions. To understand this consider the particle limit of the $3+2,$ $4+2,$
$5+2$ and $6+2$ models which was discussed in \cite{2tSuper}\cite{2ttwistor}%
. When the relativistic particle type gauge is chosen, the resulting
superparticle is described by the Lagrangian in Eq.(\ref{L1}) taken for the
corresponding dimension $d$ and the corresponding number $N$ of
supersymmetries determined by $G$. In each case, $g\left( \tau ,\sigma
\right) $ is such that it contains 32 real fermionic degrees of freedom $%
\Theta _{s}^{a}\left( \tau ,\sigma \right) ,$ but the local kappa
supersymmetry of Eq.(\ref{xi}) removes half of them so that $\theta _{\alpha
}^{a}$ in the relativistic superparticle gauge of Eq.(\ref{L1}) contains 16
real fermionic degrees of freedom. The remaining kappa supersymmetry of the
superparticle removes half of what is left, so that the \textit{physical
fermionic zero modes is 8} for each of the 3+2, 4+2, 5+2, 6+2 models. When
the superparticle is quantized in the lightcone gauge, these 8 fermionic
zero modes create $2^{3}$ bosonic physical states and $2^{3}$ fermionic
physical states, for each of the models given in the first paragraph of this
section. These states are then classified with the little group SO$\left(
d-2\right) $ in the lightcone gauge and with the $R$ symmetry group
contained in the supergroup $G$.

For $d=3,4$ we find that these superparticle quantum states are in one to
one correspondence with the physical fields of SYM theory with $N=8,4$
supersymmetries in $d=3,4$ dimensions taken the lightcone gauge  (8 bosons
and 8 fermions). There is a quick way of understanting this result. The
compactification of the $d=10$ superparticle (with its 16 fermionic degrees
of freedom) to $d=3,4$ dimensions gives the superparticle of Eq.(\ref{L1})
with the correct number of supersymmetries $N=8,4$ respectively that match
those of the gauge fixed 2T-superparticle. Thus the quantum states of the
superparticle in Eq.(\ref{L1}) must coincide with the compactification of
the physical quantum states of the $d=10$ superparticle, which is $%
8_{vector}+8_{spinor}$ of the little group SO$\left( 8\right) $ in SO$(9,1)$%
. When these SO$\left( 8\right) $ representations are reduced to the little
groups $Z_{2},$ SO$\left( 2\right) $ for $d=3,4$ respectively, they describe
the physical degrees of freedom of the superparticle as well as of SYM in $%
d=3,4$. Recall that the quantum states of the $d=10$ superparticle
correspond to the 8 bosonic and 8 fermionic fields of the 10 dimensional SYM
theory taken in the lightcone gauge. Hence the quantum states created by the
zero modes of the $3+2,$ $4+2$ string models precisely correspond to the
quantum fields of the SYM theory in the dimensions $d=3,4$ respectively.

Similarly, for $d=6$ the quantum states of the superparticle are related to
the fields of a special superconformal field theory that contains an
atisymmetric tensor $B_{\mu \nu }$ with self dual field strength $H_{\mu \nu
\lambda }=\partial _{\lbrack \lambda }B_{\mu \nu ]}=H_{\mu \nu \lambda
}^{\ast }$, five scalars $\phi ^{i}$ and fermions $\psi _{\alpha }^{a},$
with $a,i$ indicating the spinor and vector of the Sp$\left( 4\right) $ $R$%
-symmetry. In the lightcone gauge of this field theory, the transverse
degrees of freedom $B_{mn}$ describe a self dual antisymmetric tensor of the
transverse SO$\left( 4\right) =$SU$\left( 2\right) \times $SU$\left(
2\right) .$ Therefore it has 3 independent degrees of freedom classified as $%
\left( j_{1},j_{2}\right) =\left( 1,0\right) $. These together with
the 5 scalars $\phi ^{i}$ correspond to the 8 bosons, while $\psi
_{\alpha }^{a}$ supplies also 8 physical fermionic degrees of
freedom in the lightcone gauge. These are precisely the the 8 bosons
and 8 fermions produced by the superparticle of Eq.(\ref{L1}) as
follows: the gauge fixing of the superparticle all the way to the
lightcone gauge removes 3/4 of the original 32 fermions $\Theta
_{s}^{a}$ of the 2T-superparticle, leaving behind 8 zero
mode $\theta ^{\prime }s$ that are classified as $\left( \frac{1}{2}%
,0;4\right) $ under the little group SO$\left( 4\right) \times
$Sp$\left( 4\right) \subset $ SO$\left( 8^{\ast }|4\right) ,$ where
the SO$\left( 4\right) =$SU$\left( 2\right) \times $SU$\left(
2\right) $ representation is given as $\left( j_{1},j_{2}\right)
=\left( \frac{1}{2},0\right) .$ These zero modes consist of four
creation and four annihilation operators. When applied on the vacuum
they create $8$ bosons classified as $\left( 1,0;0\right) +\left(
0,0;5\right) $ and 8 fermions classified as $\left(
\frac{1}{2},0;4\right) $ under SO$\left( 4\right) \times $Sp$\left(
4\right) $. These match the transverse lightcone fields of the $d=6$
superconformal theory as described above. This special theory is
believed to be interacting and conformal at the quantum level
\cite{witConf}\cite{seiberg} but it has been difficult to study it
because of the lack of a covariant field theoretic action. The
twistor superstring formalism description given below in this paper
could be a possible approach for studying this theory in the same
way as the Witten-Berkovits twistor superstring is used to analyze
SYM theory.

The results described in the previous paragraphs give the physical spectrum
of the 2T superstring theories with the conserved current $J^{-}=\frac{1}{4}%
g^{-1}\hat{L}^{-}g$ for the superconformal groups $G$ given above.
What are the unitary representations of $G$ that emerge, and is
there a sigma model type geometrical description of these models? To
answer these questions we investigate the twistor gauge since the
same physical content of the 2T theory can be recovered in any
gauge. The result for $d=4$ is already discussed in the other
sections of this paper, while for $d=3,5,6$ it is summarized as
follows

\begin{itemize}
\item For $d=3$ the twistor $Z_{A}$ is real, it contains 8 fermions and 4
bosons, and is in the fundamental representation of OSp$\left( 8|4\right) .$
At the classical level it satisfies the condition $\bar{Z}^{A}Z_{A}=0$
automatically without constraining the degrees of freedom (OSp$\left(
8|4\right) $ metric is 1 in the fermi sector and is antisymmetric in the
bose sector). At the quantum level it contains two bosonic oscillators and
their conjugates (classified by SU$\left( 2\right) \times $U$\left( 1\right)
\subset $Sp$\left( 4\right) $) and 4 fermionic oscillators and their
conjugates (classified by SU$\left( 4\right) \times $U$\left( 1\right)
\subset $SO$\left( 8\right) $). The resulting oscillator representation in
Fock space is the supersingleton of OSp$\left( 8|4\right) ,$ and this
describes $d=3,$ $N=8$ SYM spectrum as expected from the discussion above.
In this gauge the current becomes $J_{A}^{B}=\left( \frac{1}{4}g^{-1}\Gamma
g\right) _{A}^{B}=Z_{A}\bar{Z}^{B},$ and there is a triangular subgroup $%
H_{\Gamma }\subset $OSp$\left( 8|4\right) $ that commutes with the constant $%
\Gamma ,$ whose algebra is $h_{\Gamma }=$osp$\left( 8|2\right) +V_{8|2}+R,$
such that $[osp\left( 8|2\right) ,V_{8|2}\}\sim V_{8|2}$ and $%
[V_{8|2},V_{8|2}\}\sim R,$ while the Abelian factor $R$ commutes with all
the generators in $h_{\Gamma }.$ The coset space OSp$\left( 8|4\right) /$H$%
_{\Gamma }$ has the correct counting of parameters\footnote{%
OSp$\left( 8|4\right) $ has 28+10 bosons and $4\times 8=32$ fermions. OSp$%
\left( 8|2\right) $ has 28+3 bosons and $2\times 8=16$ fermions. $V_{8|2}$
is classified as the fundamental representation of OSp$\left( 8|2\right) $
with 8 fermions and 2 bosons.} that corresponds to the twistors; these
describe the geometric space. As expected, this has the same number of phase
space degrees of freedom $\left( x,p,\theta \right) $ as the $d=3,$ $N=8$
superparticle (see footnote \ref{superdim}).

\item For $d=6$ the matrix $\Gamma $ has two non-zero entries instead of
one. Therefore there are two OSp$\left( 8^{\ast }|4\right) $ twistors $%
Z_{A}^{i},$ with $i=1,2.$ These twistors each contain 8 bosons and 4
fermions (opposite of the $d=3$ case) and are constrained as a traceless
tensor $\bar{Z}_{i}^{A}Z_{A}^{j}-\frac{1}{2}\delta _{i}^{j}\bar{Z}%
_{k}^{A}Z_{A}^{k}=0.$ These 3 bosonic constraints form an sl$\left( 2\right)
$ local symmetry$,$ therefore together with the gauge fixing of sl$\left(
2\right) $ they remove $3+3=6$ bosonic degrees of freedom from the $Z_{A}^{i}
$. Hence the number of unconstrained bosons in the geometric space described
by the twistors is $2\times 8-6=10,$ while the number of fermions is $%
2\times 4=8.$ This is the same number as the physical phase space degrees of
freedom $\left( x,p,\theta \right) $ for $d=6,$ $N=2$ superparticle (see
footnote \ref{superdim}). The coset description is found by noting that the
triangular subalgebra that commutes with $\Gamma $ is $h_{\Gamma }=$[osp$%
\left( 2,2|4\right) +$sl$\left( 2\right) $]$+V_{\left( 4|4\right) ,2}+R,$
such that $[\left( osp\left( 2,2|4\right) +sl\left( 2\right) \right)
,V_{\left( 4|4\right) ,2}\}\sim V_{\left( 4|4\right) ,2}$ and $[V_{\left(
4|4\right) ,2},V_{\left( 4|4\right) ,2}\}\sim R$ while the Abelian factor $R$
commutes with all the generators in $h_{\Gamma }.$ Thus, the geometric space
is OSp$\left( 8^{\ast }|4\right) /$H$_{\Gamma }$ with 10 bosons and 8
fermions, which is the expected number\footnote{$osp\left( 8^{\ast
}|4\right) $ has 28+10 bosons and $4\times 8=32$ fermions. osp$\left(
2,2|4\right) +sl\left( 2\right) $ has 6+10+3 bosons and $4\times 4=16$
fermions. $V_{\left( 4|4\right) ,2}$ is classified as the fundamental
representation of osp$\left( 4|4\right) $ and a doublet of sl$\left(
2\right) ,$ with $4\times 2=8$ fermions and $4\times 2=8$ bosons.}. The
quantized constrained twistors generate the oscillator representation of the
noncompact superalgebra OSp$\left( 8^{\ast }|4\right) ,$ with the
\textquotedblleft color" group sl$\left( 2\right) $ acting on the $i=1,2$
index. Only the sl$\left( 2\right) $ color singlet states are kept in the
Fock space as the physical states. The resulting representation is the
doubleton of OSp$\left( 8^{\ast }|4\right) ,$ and this indeed describes the
physical fields of the $d=6,$ $N=2,$ superconformal theory as expected from
the previous discussion above.

\item For $d=5$ we expect a geometric space F$\left( 4\right) /$H$_{\Gamma }$
whose dimension is 8 bosons and 8 fermions, since this is the counting for
the physical phase space degrees of freedom $\left( x,p,\theta \right) $ for
$d=5,$ $N=2$ superparticle (see footnote \ref{superdim}). F$\left( 4\right) $
has 24 bosons (SO$\left( 5,2\right) \times $SO$\left( 3\right) $ subgroup)
and 32 real fermions in the complex $\left( 8,2\right) $ spinor
representation of SO$\left( 5,2\right) \times $SO$\left( 3\right) .$
Therefore the triangular subgroup H$_{\Gamma }$ must contain 16 bosons and
24 fermions\footnote{$psu\left( 2|2\right) +sl\left( 2\right) $ has 3+3+3
bosons in the adjoint representation of the SU$\left( 2\right) \times $SU$%
\left( 2\right) \times $SL$\left( 2,R\right) $ subgroup, and 8 real fermions
in the complex $\left( 2,2,1\right) $ representation. $V_{\left( 3|8\right)
,2}$ is a doublet of the SL$\left( 2,R\right) $ factor and under SU$\left(
2\right) \times $SU$\left( 2\right) \subset $PSU$\left( 2|2\right) $ the
symbol $\left( 3|8\right) $ represents 3 real bosons in the representation $%
\left( 3,0\right) $ and 8 real fermions in the complex representation $%
\left( 2,2\right) .$ Therefore $V_{\left( 3|8\right) ,2}$ contains $3\times
2=6$ bosons and 8$\times 2=16$ fermions.}. Its algebra is $h_{\Gamma }=\left[
psu\left( 2|2\right) +sl\left( 2\right) \right] +V_{\left( 3|8\right) ,2}+R,$
such that $[\left( psu\left( 2|2\right) +sl\left( 2\right) \right)
,V_{\left( 3|8\right) ,2}\}\sim V_{\left( 3|8\right) ,2}$ and $[V_{\left(
3|8\right) ,2},V_{\left( 3|8\right) ,2}\}\sim R$ while the Abelian factor $R$
commutes with all the generators in $h_{\Gamma }.$ The coset space F$\left(
4\right) /$H$_{\Gamma }$ should describe a conformal theory in $d=5$ and $N=2
$ as expected from the superparticle spectrum discussed above. It is harder
to describe the space in terms of twistors because $F\left( 4\right) $ does
not have a $8|2$ dimensional fundamental representation which would have
corresponded to the spinor space of SO$\left( 5,2\right) \times $SO$\left(
3\right) .$ For the same reason the oscillator representation has not been
developed.
\end{itemize}

From the discussion above we see that each of these models can be
presented geometrically as a gauged sigma model based on the global
group $G$ and gauged with the subgroup $H_\Gamma$.

The extension from $d+2$ to higher dimensions $d+d^{\prime }+2$ with
the addition of extra $d^{\prime }$ bosons, is explained for
$4+2\rightarrow
10+2 $ with the supergroup SU$\left( 2,2|4\right) $ as in section (\ref{10+2}%
). This is slightly different in the other cases $\left( 3+2\rightarrow
11+2\right) $, $\left( 5+2\rightarrow 8+2\right) ,$ $\left( 6+2\rightarrow
11+2\right) .$ The essential difference is the kappa supersymmetry described
in Eq.(\ref{xi2}), which seems to be present only in the case $%
4+2\rightarrow 10+2$ but not in the others. The failure is due to the fact
that once the upper and lower blocks of $\hat{L}$ are normalized as $i\frac{2%
}{s}\Gamma _{MN}L^{MN}$ and $i\frac{2}{s^{\prime }}\Gamma _{IJ}L^{IJ}$ to
satisfy the local bosonic symmetry SO$\left( d,2\right) \times $SO$\left(
d^{\prime }\right) $, the kappa transformation $\delta _{\kappa }$ yields
the structure $\frac{1}{s}L^{MN}\left( \Gamma _{MN}\xi \right) +\frac{1}{%
s^{\prime }}L^{IJ}\left( \xi \Gamma _{IJ}\right) $ instead of the one that
appears in Eq.(\ref{LK}). When $s\neq s^{\prime }$ this structure does not
reduce to the SO$\left( d+d^{\prime },2\right) $ covariant dot products $%
\hat{X}\cdot \hat{X},~\hat{P}^{-}\cdot \hat{P}^{-},~\hat{P}^{-}\cdot \hat{X}$
and therefore cannot be cancelled by the variation of the Sp$\left(
2,R\right) $ gauge fields. Thus, only the case of $4+2\rightarrow 10+2$
seems to have the kappa supersymmetry given in Eq.(\ref{xi2}). Due to the
kappa supersymmetry in the case of $4+2\rightarrow 10+2,$ only 16 out of the
32 fermions in $g$ are physical degrees of freedom, and their Clifford
algebra is realized on $2^{7}$ bosons and $2^{7}$ fermions, which coincides
with the physical spectrum of type IIB supergravity in 10 dimensions. By
contrast, in the absence of kappa supersymmetry all of the 32 fermionic
degrees of freedom in the group element $g$ are physical and their quantized
zero modes give a Clifford algebra realized on quantum states consisting of $%
2^{15}$ bosons and $2^{15}$ fermions, and these may be related to the first
massive level of type IIA supergravity or the supermembrane in 11 dimensions
\cite{11Dstring}, or to the corresponding $AdS_4\times S^7$, $AdS_7\times
S^4 $ compactifications.

The reader may wonder whether other dimensions and/or supergroups may be
used in a similar fashion. This is discussed in \cite{liftM}\cite{2ttwistor}%
. An essential point to consider is the SO$\left( d,2\right) $ applied on $%
\left( X^{M},P^{M}\right) ,$ taken in the spinor representation, versus the
bosonic subgroup and the fermions in the supergroup $G.$ The supergroups $G$
of interest must contain SO$\left( d,2\right) $ in the spinor
representation, namely spin$\left( d,2\right) ,$ as a subgroup and its
fermions must be in the spinor representation of spin$\left( d,2\right) .$
In the cases discussed in this paper the spinor representation of SO$\left(
d,2\right) $ for $d=3,4,5,6$, matched precisely with one of the blocks of
the bosonic subgroup in $G,$ while the other subgroup was the R-symmetry for
$N$ supersymmetries. When this is the case the gauging of SO$\left(
d,2\right) \times $(R-symmetry) can remove all the the bosons from $G$ and
leave only the fermions in the correct spinor representation. This assures
that the remaining degrees of freedom in the particle gauge describe a
superparticle (or superstring) with usual properties. By contrast, when the
bosonic subgroups in $G$ contain more bosons than the ones in spin$\left(
d,2\right) ,$ automatically there are more physical bosonic and fermionic
degrees of freedom than just those of the superparticle or superstring. With
the type of coupling $Str\left( \partial gg^{-1}L\right) ,$ with $L$ in the
spinor representation of SO$\left( d,2\right) ,$ one finds that the extra
degrees of freedom in $g\left( \tau ,\sigma \right) $ could be related to
D-branes. Many models with brane degrees of freedom can be constructed in
this enlarged scheme. One of the most interesting cases, aiming for a
particle limit of 11 dimensional M-theory including branes, is constructed
by using the supergroup OSp$\left( 1|64\right) $ with $11+2$ dimensions
(spinor representation of SO$\left( 11,2\right) $ is 64$)$. The particle
case was briefly discussed in \cite{liftM}\cite{2ttwistor}\cite{2ttoyM} and
this is now generalized to string theory as the other cases in this paper.

As in the case of the particle, the 2T superstring in $d+2$ dimensions
discussed in this paper for several values of $d$ can be gauge fixed into a
variety of string-like 1T systems, with varying physical interpretation of
the 1T dynamics. This phenomenon has so far not been investigated in string
theory. It would be interesting to explore what one may learn about
Yang-Mills theory or string theory from these dual holographic pictures, as
well as from the unifying 2T theory that underlies them.

\begin{acknowledgments}
I would like to thank O. Andreev, N. Berkovits, M. Bianchi, C. Deliduman, M.
Vasiliev and E. Witten for discussions. I am also grateful for the
hospitality of J. Gomis at the University of Barcelona and of P.M.
Petropoulos at \'{E}cole Polytechnique, Paris, where part of this research
was performed.
\end{acknowledgments}

%\newpage

\end{document}